\documentclass[preprint]{aastex631}

\usepackage{xcolor}
\newcommand{\ME}{\rm M_{E}}

\submitjournal{PSJ}

\definecolor{nncolor}{rgb}{0.0,0.7,0}

\definecolor{rhcolor}{rgb}{1,0.5,0}

\newcommand{\revisedms}[1]{{#1}}

\shorttitle{ToF7}
\shortauthors{Nettelmann et al.}

\begin{document}

\title{Theory of Figures to the 7th order and the interiors of Jupiter and Saturn}


\author{N.~Nettelmann}
\affiliation{Dept.~of Astronomy and Astrophysics, University of California, Santa Cruz, USA}
\altaffiliation{Juno Participating Scientist}
\affiliation{Institute of Planetary Research, German Aerospace Center, 12489 Berlin, Germany}

\author{N.~Movshovitz}
\affiliation{Dept.~of Astronomy and Astrophysics, University of California, Santa Cruz, CA 95064, USA}

\author{D.~Ni}
\affiliation{State Key Laboratory of Lunar and Planetary Sciences, Macau University of Science and Technology, Macao, PR China}

\author{J.J.~Fortney}
\affiliation{Dept.~of Astronomy and Astrophysics, University of California, Santa Cruz, CA 95064, USA}

\author{E.~Galanti}
\affiliation{Department of Earth and Planetary Sciences, Weizmann Institute of Science, Rehovot, Israel 7610001}

\author{Y.~Kaspi}
\affiliation{Department of Earth and Planetary Sciences, Weizmann Institute of Science, Rehovot, Israel 7610001}

\author{R.~Helled}
\affiliation{Center for Theoretical Astrophysics and Cosmology, Institute for Computational Science, University of Zurich, CH-8057 Z{\"u}rich, Switzerland}

\author{C.R.~Mankovich}
\affiliation{Division of Geological and Planetary Sciences, Mailcode 150-21, Caltech, Pasadena, CA 91125, USA}

\author{S.~Bolton}
\affiliation{SWRI, San Antonio, TX, USA
}

\begin{abstract}
Interior modeling of Jupiter and Saturn has advanced to a state where thousands of models are  generated that cover the uncertainty space of many parameters. This approach demands a fast method of computing their gravity field and shape. Moreover, the Cassini mission at Saturn and the ongoing Juno mission delivered gravitational harmonics up to $J_{12}$.
Here, we report the expansion of the Theory of Figures, which is a fast method for gravity field and shape computation, to the 7th-order (ToF7), which allows for computation of up to $J_{14}$.  We apply three different codes to compare the accuracy using polytropic models. We apply ToF7 to Jupiter and Saturn interior models in conjunction with CMS-19 H/He-EOS.  
For Jupiter, we find that $J_{6}$ is best matched by a transition from He-depleted to He-enriched envelope at 2--2.5 Mbar. 
However, the atmospheric metallicity reaches $1\times$ solar only if the adiabat is perturbed toward lower densities, or if the surface temperature is enhanced by $\sim 14 K$  from the Galileo value.  Our Saturn models imply a largely homogeneous-in-Z envelope at 1.5--4$\times$ solar atop a small core. Perturbing the adiabat yields metallicity profiles with extended, heavy-element enriched deep interior (diffuse core) out to 0.4 $R_{\rm Sat}$, as for Jupiter.
Classical models with compact, dilute, or no core are possible as long as the deep interior is enriched in heavy-elements. Including a thermal wind fitted to the observed wind speeds, representative Jupiter and Saturn models are consistent with all observed $J_n$ values.
\end{abstract}


\section{Introduction}

Since the era of the Voyager 1 and 2 gravity field determinations and shape measurements of the outer planets, only two methods have extensively been employed to calculate the shape and the gravity field from interior models to compare with the data.  These methods are the Theory of Figures (ToF) \citep{ZT78}  and the Concentric Maclaurin Spheroids (CMS) method \citep{Hubbard12,Hubbard13}.  
ToF has served that purpose before the advent of accurate gravity data from Juno at Jupiter and from the Cassini Grand Finale Tour at Saturn. Beforehand, only the gravitational moments $J_2$, $J_4$, and $J_6$ were measured, and the smallest given uncertainty in Jupiter's $J_6$ of 10\% was still rather large \citep{Jacobson03}. 

The low-order gravitational harmonics are important observables as they constrain the density profile about midway into the planetary interior. They are expansion coefficients of the external planetary gravity field evaluated at a reference radius in the equatorial plane, $R_{\rm eq}$, which encompasses the planet's total mass. They are defined 
as integrals over density $\rho(r)$ in the planet's interior,
\begin{equation}\label{eq:Jn}
	J_{n} = -\frac{1}{M\,R_{\rm eq}^{n}}\int\! d^3r\:\rho(\vec{r})\,r^{n}\:P_{n}(\cos\vartheta)\:,
\end{equation}
where $P_{2n}$ are the Legendre polynomials, and $\vartheta$ is co-latitude.
Thanks to the Juno and Cassini missions, the observational accuracy in the even harmonics $J_{2n}$ have seen significant improvement \citep{Iess18,Durante20,Iess19}. 
For both Jupiter and Saturn, the uncertainties in the low-order harmonics reduced to a level that can be considered exact from the perspective of adjusting internal density distributions to reproduce the data. However,  significant spread in the deep interior density distributions is still possible, see \citet{Movshovitz20} for Saturn,  as the sensitivity of the $J_{2n}$ toward the center fades with $(r/R_{eq})^{2n}$, see Eq.~(\ref{eq:Jn}). This spread is a residual uncertainty related to other causes such as the positioning of internal helium gradients  due to uncertainty in the H-He phase diagram, the temperature  profile in stably stratified regions, the H-He equations of state, or the positioning of heavy element gradients due to uncertainties in planet formation and evolution.

ToF to the fourth order (ToF4) has been deemed sufficiently accurate for computation of $J_2$ and $J_4$ usually used to constrain the density distribution \citep{Nettelmann17}.
Higher order moments beyond $J_4$ have been provided to a precision in $J_6$ to $J_{10}$ of better than (0.01--0.1)$\times 10^{-6}$ for Jupiter \citep{Durante20} and  (0.1--1)$\times 10^{-6}$ for Saturn \citep{Iess19}, but as the order increases so does the influence by the zonal flows on the harmonics. At present, this is where the limitations of the ToF method become evident. ToF is an expansion method. An $n$th-order expansion (ToF$n$) allows to compute up to $J_{2n}$ and to an error of the order of $q_{\rm rot}^{n+1}$, where $q_{\rm rot}=\omega^2R_{eq}^3/GM$ is the ratio of centrifugal to gravitational force at the equatorial radius. The highest presented ToF-order so far is 5 \citep{ZT75}. It has recently been applied to compute Saturn's $J_2$--$J_{10}$ values \citep{Ni20}, however, its accuracy has not been validated yet.

Although the CMS method is also an expansion method, it can conveniently be carried out to the order of 15-20 or higher \citep{Hubbard13}. Therefore, it enables high-accuracy computation of the high-order $J_{2n}$ up to the order of the measurements \citep{Wahl17J,Militzer19}. The CMS method provides further advantages such as its expansion to 3D to account for tidal shape and gravity field perturbations \citep{Wahl17S} and brevity in its formulation \citep{Hubbard13}.
Its only drawback is that CMS method goes along with high computational cost even in its accelerated version \citep{Militzer19}. This is because CMS method explicitly solves for the 2D planetary shape not only taking the sum over radial spheroids but also by integrating over latitude. Even if making use of Gaussian quadrature, typically several tens of angular grid points are required. If the integrand were a polynomial, only $N_{\rm lat}=n+1$ grid points would be required to evaluate the integrals over $P_{2n}$, which for $J_{12}$  $(n=6)$ amounts to only $N_{\rm lat}=7$, or even 4 points when accounting for hemispheric symmetry. However, the integrands are functions of the non-polynomial shape itself. In practice, 48 grid points \citep{Wahl17S} are used. Obtaining the shape to sufficient accuracy at these grid points is the most time-consuming part of the CMS method. In contrast, ToF solves for the shape explicitly only in the equatorial plane while the shape at higher latitudes is obtained by spherical harmonics expansion, and the required precision of the shape is only the one wanted for the $J_{2n}$. One may thus see a benefit in using ToF for computation of the high-order moments.

Here, we introduce ToF7 tables, which allow one to calculate up to $J_{14}$. In Section \ref{sec:methods}, we \revisedms{give an overview of the ToF method while for further details we refer to Appendix \ref{sec:app_ToFnotes}.} In Section \ref{sec:poly}, we assess the accuracy of the ToF method by comparing to the analytic $n=1$ polytrope solution. In Section \ref{sec:Jupiter}, we apply the new tables  to Jupiter models, and in Section \ref{sec:Saturn}  to Saturn. In Section \ref{sec:winds}, we connect representative interior models to thermal wind models to predict the wind decay depth profiles. 
\revisedms{Observing notoriously low atmospheric metallicities of our Jupiter models, we discuss further influences in Section \ref{sec:discus}. Section \ref{sec:conclusions} concludes the main body of the paper. In Appendix \ref{sec:app_tables} we introduce the ToF7 tables for public usage}.

\section{Theory of Figures}\label{sec:methods}

The Theory of Figures is described in \citet{ZT78} and the coefficients up to the 3rd order presented therein. \citet{Nettelmann17} followed their notation and calculated the 4th-order coefficients. We note that 5th-order coefficients were presented in \citet{ZT75} and adopted by \citet{Ni20} for application to Saturn. Building upon the work of \citet{Nettelmann17}, we here conduct the expansion of ToF to the 7th-order, meaning that the even harmonics up to $J_{14}$ can be calculated. 

Both the ToF and CMS methods assume that surfaces of equal potential $U$ exist on which density and pressure are constant. One can show that this assumption holds for planets in hydrostatic equilibrium that rotate along cylinders, e.g., if their rotation rate can be expressed as $\vec{\omega} = \omega(s)\:\vec{e}_{\omega}$ with axis distance $s$. Rigid rotation and no rotation meet this condition. For rotation along cylinders, the odd harmonics $J_{2n+1}$ disappear. However, the Juno measurements at Jupiter \citep{Iess18,Durante20} and Cassini Grand Finale at Saturn \citep{Iess19}  revealed that these planets' odd harmonics are non-zero. Instead, they are of the order of $0.1\times 10^6$, comparable to the values of $J_{10}$, $J_{12}$, and Saturn's uncertainty in $J_6$. Based on the commonly used approach of using the thermal wind equation (TWE) to infer the density anomalies \citep{Kaspi10, Kaspi13}, the depth of the wind-induced deviation from cylinder-rotation has been inferred to be about 3000 km ($\sim 0.035\:R_{\rm Jup}$) in Jupiter \citep{Kaspi18} and 9000 km ($\sim 0.14 R_{\rm Sat}$) in Saturn \citep{Galanti19}. These results are consistent with the tangent-cylinder model of \citet{Dietrich21}, which goes beyond the TWE simplification by including not only the wind-induced density perturbation but also the associated gravitational perturbation, what has long been argued to be significant \citep{Zhang15}. Taking into account also constraints from the observed secular variation of the magnetic field on deep flows \citep{Moore19}, suggests a somewhat steeper decay function for the winds, with the zonal flow extending inward on cyclinders almost barotropicaly to a depth of about 2000 km on Jupiter and 8000 km on Saturn and then the winds decay abruptly within then next ~1000 km \citep{GalantiKaspi21}.  While the non-asymmetric gravity field is important for these width depth issues, here we have to neglect the asymmetries, as otherwise neither the ToF nor the CMS method could be applied.

In the absence of  tides, the problem at hand is axisymmetric and thus 2D: $r,\vartheta$. In ToF, the description is further reduced to 1D by introducing the mean radius coordinate $l$. Spheres of radius $l$ are defined by the condition that they enclose the same volume as the equipotential surface $r_l(\vartheta)$,
\begin{equation}\label{eq:equalvoll}
	\frac{4\pi}{3}\: l^3 = 2\pi\int_0^{\pi} \!\sin\vartheta d\vartheta \int_0^{r_l(\vartheta)}\! \:dr'\! r'^2 \:.
\end{equation}
On the surface of the planet,
\[
	\frac{4\pi}{3}\: R_m^3 = 2\pi\int_0^{\pi} \!\sin\vartheta d\vartheta \int_0^{R(\vartheta)}\! \:dr'\! r'^2
\] 
with $R(\pi/2) = R_{eq}$. In ToF, the potential is thus constant on spheres. Both the total potential $U(l)$  and the axisymmetric 2D-shape $r_l(\vartheta)$ are expanded into Legendre polynomials. One can write 
\begin {equation}\label{eq:U_Ak}
	U(l)=\frac{GM}{R_{m}}\:\left(\frac{l}{R_m}\right)^2\sum_{k=0}^O A_{2k}(l)\:P_{2k}(\cos\vartheta)\:,
\end{equation}
while the shape of an equipotential surface, also called level surface, is given by
\begin{equation}\label{eq:rl}
	r_l(\vartheta) = l\left(1+\sum_{k=0}^O s_{sk}(l)\:P_{2k}(\cos\vartheta) \right)\:,
\end{equation}
where the $s_{2k}(l)$ are the figure functions. The condition that $U(l)$ is constant on spheres of radius $l$ implies that $A_{2k}=0$ for $k>0$. These expansions are carried out up to an order $O$. 
In the absence of tides, $U$ is a superposition of only the gravitational potential $V$ and the centrifugal potential $Q$ so that $U=V+Q$ and $A_{2k} = A_{2k}^{(V)} + A_{2k}^{(Q)}$.
By definition, the gravitational harmonics $J_{2n}$ can be obtained in the ToF  as 
\begin{equation}\label{eq:DN_JiS}
	J_{2i} = -(R_m/R_{eq})^{2i} S_{2i} (1) \:,
\end{equation}
where the integrals $S_{2i}$, not to be confused with the figure functions $s_{2i}$, are given by Eq.~(\ref{eq:Snfn}) in the Appendix.
However, using Eq.~(\ref{eq:Snfn}) for the $S_n$ and the ToF-expansion coefficients to calculate the functions $f_n$ on which the $S_n$ depend, see Eqs.~(\ref{eq:Snfn},\ref{eq:fn}) in the Appendix, implies that only information on the equatorial radius $r_l(\pi/2)$ enters the computation of the $J_{2n}$, while information on the full shape $r{(l,\theta)}$ is reduced to the order of the expansion, that is up to $P_{14}$. An alternative method is to calculate the integrals over latitude explicitly. In Section \ref{sec:polyDN} we compare both methods.
\revisedms{For details on how the ToF-coefficients are computed and for an example of the machine-readable asci-tables that contain their values for public usage, see Appendix \ref{sec:app_tables}.}  \revisedms{Moreover, to facilitate the application of our ToF7 tables, we share computer routines for read-in of the tables and documentation at {\tt https://doi.org/10.6084/m9.figshare.16822252. }}

\section{Validation against the \textit{n} = 1  polytrope}\label{sec:poly}

The $n=1$ polytropic planet is specified by a number of conditions. First, the polytropic EOS $P=K\:\rho^2$. Furthermore, the gravity field of the rotating polytrope depends on the values of $q_{\rm rot}$, equatorial radius $R_{eq}$, and planet mass $M$. The density profile $\rho(r)$ is not known in advance but is obtained from solution of the equation of hydrostatic equilibrium, $\nabla P/\rho = \nabla U$. In ToF, the radial coordinate  is taken the level surface $l$, and the \revisedms{equation of hydrostatic equilibrium} reduces to $dP/dl = \rho dU(l)$.
The internal $m$--$l$ relation is obtained by integrating the equation of mass conservation,  $m(l) = 4\pi\int_0^l dl\:\rho(l) l^2$. The latter is a source of numerical inaccuracy. We employ three different codes to compute the solution to the rotating polytrope. All polytropic models use $q_{\rm rot}=0.089195487$ and $GM=126686536.1\times10^9$ km$^3/$s as in \citet{WH16}. 

Before we compare the results of our application of three different codes and different orders of expansion of ToF to the analytic,  Bessel-functions based method of \citet{WH16} in Figure \ref{fig:poly}, we describe each of the three employed methods in Sections \ref{sec:polyNN}--\ref{sec:polyDN}.

\subsection{Polytrope with \textsc{Mogrop}}\label{sec:polyNN}

In the \textsc{Mogrop} code \citep{Nettelmann17}, the constant $K$ is adjusted to fit the mass $M$. The mean radius $R_m$ is adjusted to fit $R_{eq}$. The radial grid, for this application, is split into $N$ grid points,  out of which $N/2$ are equally distributed over 0--0.95 $R_m$, and the other half equally over 0.95--1 $R_m$. 
Such a choice was found to give a better match to the analytic solution than a split at 0.9 $R_m$ or deeper. Indeed with MOGROP, 
we find that the accuracy increases the farther out the separation is made, with a difference up top an order of magnitude 
compared to a flat distribution. The integrals in Eqs.~(\ref{eq:Snfn},\ref{eq:Snpfnp}) are converted into integrals over 
density by partial integration and solved by simple trapezoidal rule. The integration of the equations of mass conservation, 
$dm/dl=4\pi\,l^2\rho(l)$, and hydrostatic equilibrium, $(1/\rho(l))\,dP/dl=dU/dl$, is performed by the RungeKutta 4th-order method. 
The $J_{2n}$ are computed using Eq.~(\ref{eq:DN_JiS}) and denoted by '{4,7/Ne}' and shown as green curves in Figure \ref{fig:poly}.

\subsection{Polytrope with \textsc{TOF-planet}} \label{sec:polyNM}

The second code we use in our $n=1$ polytrope comparison test case is an independent implementation of the ToF algorithm using the same coefficients but otherwise unrelated to the \textsc{Mogrop} code. The two codes are therefore expected to reproduce very similar solutions if given the same conditions. \textsc{TOF-planet} has previously been applied in a baesian study of Saturn's possible interior \citep{Movshovitz20}. Since for that purpose it was necessary to run tens of millions of density models to draw representative statistical samples, the code had to be optimized for speed and memory usage. An optional feature allows the shape functions to be explicitly calculated on a subset of level surfaces, while the shape of the rest can be spline-interpolated in the radial direction. This ``skip-n-spline'' trick can provide a significant speed advantage when high resolution density profiles are needed. We find that, even when a very high resolution of the density profile is required to accurately calculate integrals over density, there is no advantage in calculating the shape functions for more than a few hundred level surfaces.
\revisedms{
 The speed advantage of this optimization applies mainly to high-resolution ToF7 calculations. For lower resolution, and for most ToF4 runs the interpolation overhead ruins the effort. (ToF7 is much slower than ToF4 for a given $N$ owing to the many more terms appearing in each of the shape function equations.) 
}

To validate \textsc{TOF-planet} with both ToF4 and ToF7 coefficients, we use it to reproduce the $n=1$ polytrope test of \citet{WH16}. To make a direct comparison in a consistent way, some care is needed. The mass and equatorial radius are taken as in \citet{WH16} and remain fixed for the duration of the calculation. (The mass is taken from the reported $GM$ and with $G=6.6738480\times{10}^{-11}\;\mathrm{m^3\;kg^{-1}\;s^{-2}}$). However, in \citet{WH16} the rotation state is given by the parameter $q_\mathrm{rot}$ whereas the ToF algorithm needs the related parameter $m_\mathrm{rot}$. The conversion needs the ratio $R_\mathrm{eq}/R_\mathrm{m}$, which is only available after the equilibrium shape is solved. To obtain a self-consistent solution we fix the planet's rotation frequency $\omega$ using the value $q_\mathrm{rot}=0.089195487$. We then proceed with a guess for $R_\mathrm{eq}/R_\mathrm{m}$ and therefore $m_\mathrm{rot}$, solve for the shape function and gravity field, integrate the hydrostatic equilibrium equation to solve for pressure everywhere, update the density everywhere to match the polytropic relation, renormalize the level radii grid to match the reference equatorial radius, renormalize the density to match the reference mass, recalculate $m_\mathrm{rot}$ for the updated $R_\mathrm{eq}/R_\mathrm{mean}$ ratio, and rerun all the steps until a self consistent solution is found.

 \revisedms{
 In this test, for both ToF7 and ToF4, we compute all integrals with the trapezoidal rule and constant grid spacing. With TOF-planet, we experimented and found that different integration schemes and grid spacing schemes did not reduce substantially the number of grid points required for a given precision.
 This should not discourage, however, future users of our ToF7 tables from optimizing their grids and integration schemes for their particular cases. 
}
The resultant $J_{2n}$ values appear in blue and are denoted by '{4,7/Mo}' in Figure \ref{fig:poly}.

\subsection{Polytrope with CEPAM} \label{sec:polyDN}

As in previous work \citep{Ni20} we apply the CEPAM code \citep{GuillotMorel95} to calculate the gravity field and shape using ToF5 \citep{ZT75}. Here, we have expanded the code to address the case of the rotating $n=1$ polytrope.
 
For the $n=1$ polytropic EOS $P = K\rho^2$, the constant $K$ is determined in terms of mass conservation and the mean radius $R_m$  is adjusted to reproduce the equatorial radius $R_{\rm eq} = 71492$ km.  The initial density distribution is firstly given by that of a nonrotating $n=1$ polytrope $\rho(z) = \rho_c \sin \pi z/\pi z$. The figure function $s_{2k}(z)$ and total  potential $U(z)$ are computed using the ToF5 as described in \citet{ZT75} and \citet{Ni19}. 
\revisedms{In its original version, CEPAM uses an automatic grid refinement method that distributes the grid points in a way that a distribution function of the variables pressure, temperature, mass, radius, and luminosity changes by a constant amount at the grid points. This method requires smooth behavior of the variables and their derivatives. B-splines are used as the interpolating polynomials, which exhibit the desired properties. However, for number of grid points larger than $10^3$, we did not obtain stable solutions with CEPAM. Therefore, for higher number of grid points we switched to our own solution of the  pressure profile using the trapozoidal rule}
\begin{equation}
	P(z_j) = P(z_{j-1}) + 0.5[\rho(z_j) + \rho(z_{j-1})] [U(z_j) - U(z_{j-1})]
\end{equation}
with the outer boundary  condition $P(R_m)=0$ Mbar. \revisedms{In this case and} in view of the fact that gravitational harmonics show greater sensitivities to the external levels of a planet, more radial grid points are taken for the outer part: $N/2$ equally distributed over 0.85--1 $R_m$ and the other half equally over 0--0.85 $R_m$. \revisedms{With CEPAM, we find this choice yields a modest optimum in accuracy.}

Finally, the new density distribution is obtained from the $n=1$ polytropic EOS $\rho(z_j) = \sqrt{P({z_j})/K}$. 
This procedure is iteratively performed until all changes in the density distribution are reduced to  within  a specified tolerance.\\
In the work of \citet{Ni20}, the gravitational zonal harmonics are calculated as weighted integrals over the internal density distribution $\rho(z)$ using the resulting figure functions $s_{2k}(z)$,
\begin{equation}\label{eq:DN_JiT}
	J_{2i} = -\frac{2\pi}{M R^{2i}_{eq}} \int^{+1}_{-1} d\cos\theta P_{2i}(\cos\theta)T (R_m,\theta) ,
\end{equation}
\begin{equation}
	T(R_m, \theta) = \int^{R_m}_0 \!dl\: \rho(l, \theta)r^{2i+2} \Big(\frac{dr}{dl}\Big)\:.
\end{equation}
Using the scaled mean radius $z=l/R_m$ and abbreviating the level surface Eq.~(\ref{eq:rl}) as  $r=z R_m[1 + \Sigma(z,\theta)]$, one can express the function $T(R_m, \theta)$ as
\begin{equation}
	T(R_m, \theta) = \frac{3M\, R^{2i}_m}{4\pi}\left\{ \frac{ [1 + \Sigma(1,\theta)]^{2i+3}}{2i+3}
	\: \frac{\rho(1,\theta)}{\bar{\rho}} - \int_0^1 \!dz\: 
	\frac{[1 + \Sigma(z, \theta)]^{2i+3}}{2i + 3}\,z^{2i+3} \: \frac{d\rho(z, \theta)/\bar{\rho}}{dz}\right\}.
\end{equation}

\begin{table}
\centering
\vspace*{0.2cm}
\caption{\label{tab:polyDN}
Comparison of the $J_{2n}$ obtained from Eq.~(\ref{eq:DN_JiT}) and from Eq.~(\ref{eq:DN_JiS})} 
\begin{tabular}{cccccc}\hline\hline
Method  & $J_2 \times 10^6$ & $J_4\times10^6$  & $J_6\times 10^6$  & $J_8\times 10^6$  & $J_{10}\times10^6$ \\\hline
Bessel  &  13988.51 & -531.8281 & 30.11832 & -2.13212 & 0.17407\\
Eq. (\ref{eq:DN_JiT})  & 13988.54  & -531.8292 & 30.11989 & -2.13048  &0.17446\\
$|\Delta J_{2i}/J^{\rm Bessel}_{2i}|$ & $2.42\times10^{-6}$ & $1.99 \times 10^{-6}$ & $5.22\times 10^{-5}$ 
& $7.67\times 10^{-4}$ & $2.23\times 10^{-3}$\\
Eq. (\ref{eq:DN_JiS}) & 13988.55 & -531.8207 & 30.13506 &-2.10486 &0.19555\\
$|\Delta J_{2i}/J^{\rm Bessel}_{2i}|$ & $2.52\times10^{-6}$ & $1.40 \times 10^{-5} $ & $5.56\times 10^{-4}$ 
& $1.28\times 10^{-2}$ & $1.23\times 10^{-1}$\\
\hline\hline
\end{tabular}
\begin{minipage}{0.8\textwidth}
\vspace*{0.2cm}
\textsc{Note}---The numerical values in this Table are for a typical number of grid points $N = 2000$ and using the ToF5 coefficients of \citet{ZT75}. The Bessel-solution is taken from \citet{WH16}.
\end{minipage}
\end{table}
Table \ref{tab:polyDN} shows a comparison of the even harmonics obtained from Eq.~(\ref{eq:DN_JiT}) and from Eq.~(\ref{eq:DN_JiS}) for a typical number of grid points $N=2000$. The numerical accuracy in $J_2$  is almost the same for both of them. But the results from Eq.~(\ref{eq:DN_JiT}) are in better agreement with the analytic Bessel-functions based solution for $J_4$--$J_{10}$, where the numerical accuracy for Eq.~(\ref{eq:DN_JiT}) is about 1--2 orders of magnitude better than that for Eq.~(\ref{eq:DN_JiS}).

\subsection{Comparison of the $J_{2n}$ of the uniformly rotating polytrope}

\begin{figure*}
\centering
\includegraphics[width=0.98\textwidth]{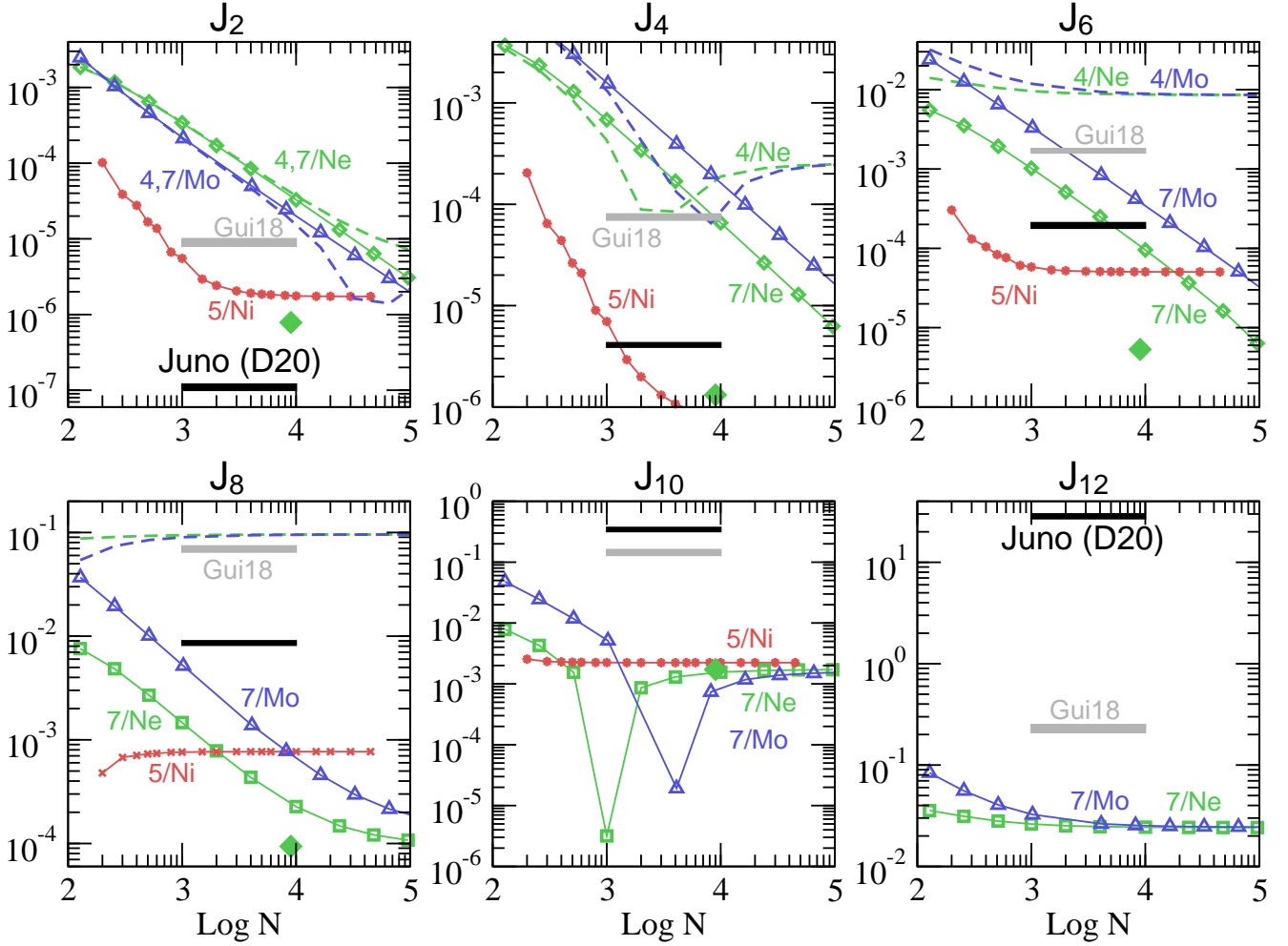}
\caption{\label{fig:poly}
Relative differences $|\Delta J_{n}/J_n|$ between ToF solutions and the analytic Bessel-functions based solution \citep{WH16}.
Dashed lines (green and blue): ToF4;
lines with open symbols (green and blue): ToF7;
red lines with stars: ToF5 using CEPAM and labeled 5/Ni,
green lines and labeled 4/Ne or 7/Ne: using \textsc{Mogrop}, 
blue lines and labeled 4/Mo or 7/Mo: using TOF-planet;
grey bars labeled Gui18: CEPAM-WH16 from \citet{Guillot18},
black bars labeled Juno (Du20): obs.~uncertainty \citep{Durante20}. 
X-axis shows number $N$ of radial grid points used in this work.
\revisedms{Green diamonds placed at $N=9000$ are for extrapolated $J_{2n}$ values based on linear regression on the three computed values at $N=1000$, 2000, and 4000.}}
\end{figure*}

In Figure \ref{fig:poly} we show the relative deviations of the calculated even harmonics from the analytic solutions of \citet{WH16} as a function of the number $N$ of radial grid points.

With CEPAM code and ToF5, denoted by 5/Ni in the Figure, the numerical accuracy of all the calculated $J_{2i}$ shows good convergence with an increased number of grid points. When the number of grid points is increased beyond $\sim 10^3$, the numerical accuracy in $J_4$--$J_{10}$ falls below the current observational  uncertainty (Juno D20) reported in \citet{Durante20}. Moreover, the accuracy in all the harmonics $J_2$--$J_{10}$ is better than the CEPAM-WH16 results from \citet{Guillot18}, who reportedly applied ToF4, by a factor of roughly 5--100.

Using ToF4 and ToF7 in conjunction with the \textsc{Mogrop} code, denoted by 4/Ne and 7/Ne in the Figure, the accuracy significantly improves with denser grid points. Apparently, this code requires a factor of 100 more radial grid points than CEPAM to obtain the same accuracy in $J_2$ and $J_4$. For these low-order harmonics, ToF7 vs.~ToF5 provides a negligible improvement in accuracy. The situation changes with $J_6$. Here, CEPAM with ToF5 levels off at a relative  uncertainty of $\sim 5\times 10^{-5}$, while the higher accuracy of ToF7 vs.~ToF5 becomes evident as N increases beyond 20,000. However, the typical number of grid points used for planet interior models ranges between $2000$ and 4000. 
For such $N$ values, the numerical accuracy in $J_6$ with 7/Ne  is about the same as the current observational uncertainty reported in \citet{Durante20}.  ToF7 begins to pay off with $J_8$ and higher, even with \textsc{Mogrop}, where $J_8$ becomes an order of magnitude better than 
the observational one, 2.5 orders of mag in $J_{10}$, and 3 orders of mag in $J_{12}$. We conclude that ToF7 is sufficiently accurate to address the influence of the winds on $J_6$ and higher, given current observational uncertainties.

Using the independent \textsc{TOF-planet} code of \citet{Movshovitz20}, we obtained similar $J_{2n}$ values as with the \textsc{Mogrop} code, compare the blue and the green lines in Figure \ref{fig:poly}. In particular, the results for $J_4$--$J_8$ with ToF4 
after convergence with grid point number $N$ agree, indicating that the remaining errors $\Delta J_n/J_n$ of $2.5\times 10^{-4}$ in $J_4$, $10^{-2}$ in $J_6$, and $10^{-1}$ in $J_8$ are due to the truncation of the expansion of ToF4. The ToF7 values also agree when convergence is reached, though this applies only to $J_{10}$ and $J_{12}$ for high values of $N>10,000$. Before convergence with $N$ is reached, the ToF7 errors deviate by a factor of a few, suggesting that the radial grid discretization error matters, which can differ between different implementations even if they use the same trapezoidal rule. 

\revisedms{The similar accuracy of \textsc{Mogrop} and TOF-planet is due to similar methods for the numerical integration over density (trapezoidal rule) and of the differential equations $dm/dr$ and $dP/dr$ (Runge-Kutta). There is room for improvement. As an example, we extrapolate the $J_{2n}$ values using linear regression on the three solutions for $N=1000$, $N=2000$, and $N=4000$ for each $J_{2n}$. In Figure \ref{fig:poly}, the resulting accuracy is conservatively compared to the result for $N=9000$ corresponding to a the computational cost that scales linearly with $N$ and a small offset for each run.}  \revisedms{Apparently, the gain in accuracy amounts two orders of magnitude in $J_2$ and $J_4$, and is still better than compared to using $10^5$ grid points. This suggests that methods other than simply sky-rocketing the number of grid points may help to improve  the accuracy of numerical $J_{2n}$ computations.} 

We note that \revisedms{at present}, ToF is entirely outperformed by the CMS method in regard to accuracy,. \citet{Militzer19} reported relative inaccuracies of only $7.3\times 10^{-9}$ in $J_2$, $2.1\times 10^{-10}$ in $J_4$, $3.6\times 10^{-8}$ in $J_6$, $4.2\times 10^{-8}$ in $J_8$, $1.1\times 10^{-7}$ in $J_{10}$, and
$6.7\times 10^{-9}$ in $J_{12}$ for $N=2^{17}=131072$ CMS layers, out of which only 512 are treated explicitly, while the shape of intermediate ones is obtained by interpolation.

\section{Application to Jupiter}\label{sec:Jupiter}

\subsection{Jupiter models}\label{sec:Jmodels}

The models of this work assume a four-layer structure. By $Y_i$ we denote the helium mass fraction in layer number $i$ with respect to the H/He system. Layer 4 is a compact rocky core. Layer 1 is an atmosphere with a helium mass fraction of  $Y_1=0.238$ as measured by the Galileo entry probe. Layers 2 and 3 have the same helium abundance ($Y_2=Y_3$), which is adjusted to yield an overall helium mass fraction $Y=0.2700(4)$. A possible He-rain region in Jupiter is represented by a jump in  helium abundance between layers 1 and 2. Transition pressures of 1--4 Mbar for $P_{12}$ are considered, which are typical pressures where the Jupiter adiabat reaches the closest point to the H/He-demixing boundary of H/He phase diagrams for protosolar H/He ratios as predicted by  first-principles simulations \citep{Lorenzen11, Morales13, HM16, Schoettler18}. Adjusting  the local He abundance to the local $P$-$T$ conditions along the phase boundary yields an approximately linear increase in $Y$; however, the gradient and width of the He-rain region depends on the temperature profile assumed therein \citep{Nettelmann15}, which may range  in Jupiter from adiabatic to modest superadiabaticity \citep{Mankovich20}.
A recent analysis of reflectivity data obtained for H/He samples that were pre-compressed to 2--4 GPa in Diamond Anvil Cells and further shock-compressed to 60--180 GPa using the OMEGA layer indicate that an even larger portion of the Jupiter adiabat may intersect with the H/He phase boundary, as at the highest pressure where evidence of demixing is seen, 150 GPa, the measured temperatures were 10,000 K \citep{Brygoo21}. Assuming a flat $T(P)$ phase curve at Mbar pressures, such a temperature corresponds to $\sim 8$ Mbar along the Jupiter adiabat \citep{HM16}.

Although He droplets may carry specific elements such as Ne with them downward \citep{WiMi10} and affect the metallicity between the He-depleted outer and He-enriched inner region, we assume constant heavy element mass fractions
across that boundary ($Z_1=Z_2$). Finally, between layers 2 and 3, the heavy element mass fraction is allowed to change. We either use a constant $Z_3$-value implying a jump in $Z$ at a transition pressure $P_{23}$, or a Gaussian $Z_3$-profile that starts with $Z(P_{23})=Z_2$ and smoothly increases toward a maximum $Z_{3,max}$ at $P=38$ Mbar near the core. The choice of 38~Mbar is arbitrary and was taken to be just atop usual core-mantle-boundary pressures, which are found to be around 40 Mbar in Jupiter.
The two free parameters in that setup to adjust $J_2$ and $J_4$ are $Z_1$ and $Z_3$ or $Z_{3,max}$. 

We employ the CMS-2019 equations of state (EOS) for H and He \citep{Chabrier19} and mix them with the water EOS H2O-REOS with respect to density only. The $T$--$P$ profile is that of the H/He adiabat, which begins at T=166.1 K at 1 bar. We construct curves of constant entropy (adiabats) by using the specific entropy values $s_{\rm H}(P,T)$ and $s_{\rm He}(P,T)$ provided in the tables for hydrogen and helium \citep{Chabrier19} after adding a composition-dependent entropy of mixing term $s_{\rm mix}(X_{\rm H_2},X_{\rm He})$, For the concentrations, we assume that helium is non-ionized and that hydrogen is either molecular or ionized, taking the degree of ionization as in \citet{Nettelmann08}. Since we found these H/He adiabats to be too dense to yield Jupiter models with non-negative atmospheric metallicity, we also perturb that adiabat toward lower densities as described in Section \ref{sec:Jup_rho}.

\subsection{Results for Jupiter's even harmonics}

\begin{figure*}
\centering
\includegraphics[width=0.84\textwidth]{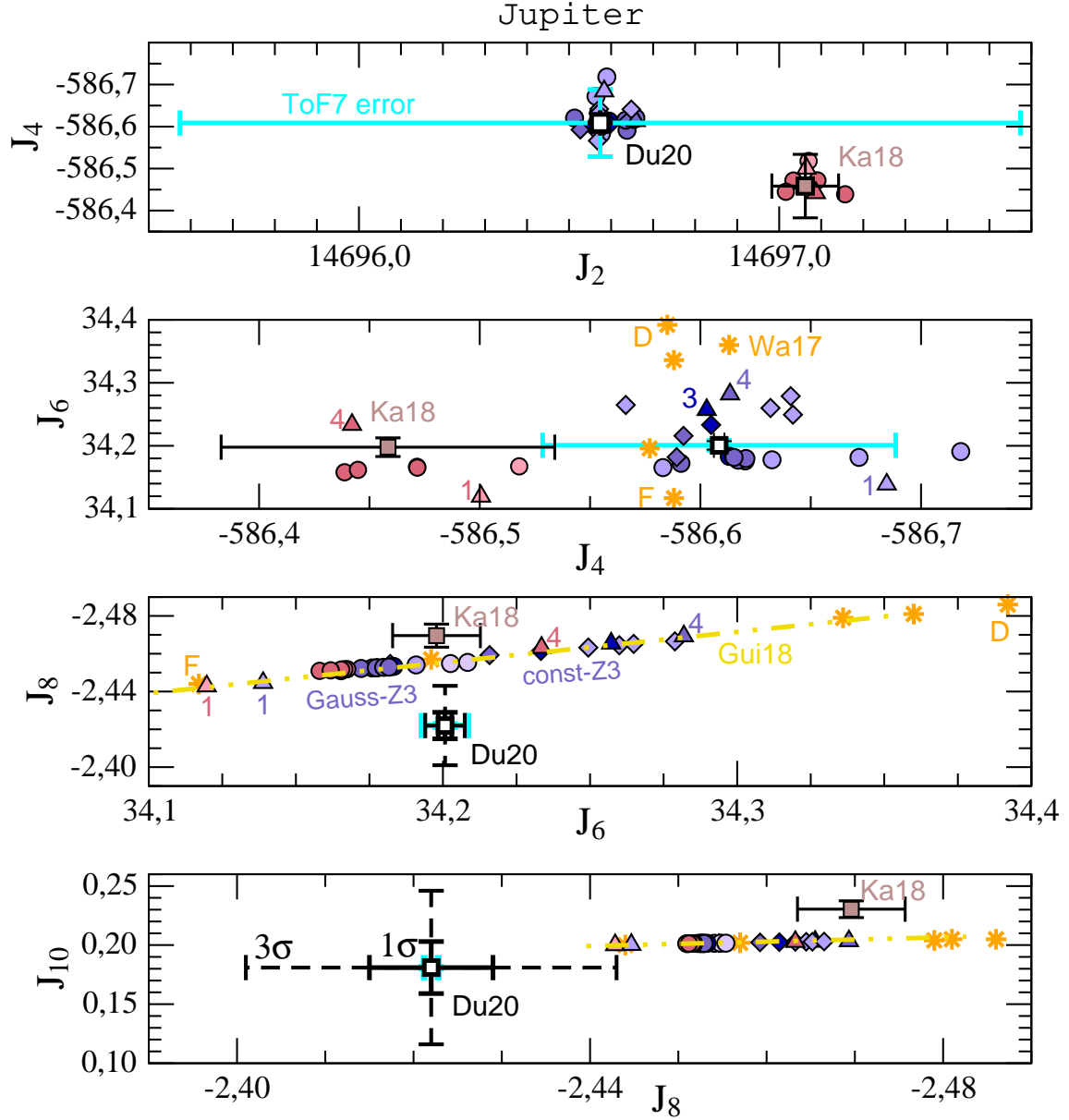}
\caption{\label{fig:jupJ2n}
$J_{2n}$ values  observed by Juno (white squares), corrected for latitude-dependent winds (brown squares; \citealp{Kaspi18}),  models with MH13 EOS (orange stars, \citealp{Wahl17J}), models in Extended Data Figure 1 of \citet{Guillot18} (yellow), models with CMS19 H/He-Eos and $P_{12}=1,3,4$ Mbar (triangles), $P_{12}=2$ Mbar and  $P_{23}$ varied from 5 to 20 Mbar and constant $Z_3$ (diamonds) or Gaussian $Z_3$ (circles). \revisedms{All $J_{2n}$ values are multiplied by $10^6$.}}
\end{figure*}

In  Figure \ref{fig:jupJ2n} we show the even $J_2$--$J_{10}$ values from rigidly rotating Jupiter models. Models adjusted to the Juno-observations of $J_2$ and $J_4$ \citep{Durante20} are shown in bluish color, while models adjusted to the wind-corrected $J_2$ and $J_4$ values by the corrections of \citet{Kaspi18} applied to the $J_2$, $J_4$ values of \citet{Durante20} are shown in reddish color.

Due to imperfect fit to the $J_2$, $J_4$ values, the scatter in model $J_2$ and $J_4$ values is larger than the observational uncertainty. For $J_4$, the scatter $\Delta J_4/J_4$ is about $\pm 2\times 10^{-4}$ and of same size as the relative uncertainty due to using ToF7 in the \textsc{Mogrop} code, while for $J_2$ the latter relative uncertainty is with $7\times 10^{-5}$ overwhelming. Still, these relative deviations are too small 
to matter for the inferred metallicities.  \citet{Guillot18} allowed for a similarly wide scatter in $J_2$ model values of $\pm 3.4\times 10^{-5}$ and a wider scatter in $J_4$ of $\pm 10^{-3}$ relative deviations. They found that nevertheless, the high-order moments 
$J_8$ vs.~$J_6$ and $J_{10}$ vs.~$J_8$ were strictly confined to a straight line. We confirm that behavior.

Notably, the model $|J_8|$ values are higher than the observed value, and a trend in that direction is also seen for $J_{10}$ although the model $J_{10}$ values are still within the $3\sigma$ observational uncertainty. Wind models assuming rotation along cylinders indeed predict a slight decrease in $|J_8|$ and $J_{10}$ if the observed wind profile of the southern hemisphere is applied to the entire surface, while they  predict an enhancement if  the wind profile of the northern hemisphere is used \citep{Hubbard99}. The wind model by \citet{Kaspi18} that was adjusted to explain the odd moments of Jupiter observed by 
Juno yields a correction qualitatively in the direction as predicted for the southern winds and seen in the model values for rigid rotation, albeit quantitatively stronger by a factor of two. This deviation may have many reasons; clearly, further exploration of the connection between interior and wind models is desirable.

For $J_6$, the uncertainties from the application of ToF7 in the \textsc{Mogrop} code, from observations, and from the wind contribution are all small and of same size. In contrast, model assumptions such as the location of layer boundaries not only have a larger influence on $J_6$ but also yield a scatter \emph{around} the observed value. Hence, we conclude that Jupiter's $J_6$ is unique in that it is neither adjusted nor seems to be significantly influenced by the winds,  and therefore offers an additional parameter to further constrain interior models. 
We find that models with a Gaussian-$Z_3$ and an abrupt He-poor/He-rich transition at $P_{12}=$1--2 Mbar yield $J_6=34.11$--34.18 slightly below the observed value, while models with that transition deeper inside at $P_{12}=$2.5--3 Mbar yield $J_6=34.23$--34.28 slightly above the observed value. Constant-$Z_3$ profiles and $P_{12}=$2 Mbar stretch from $J_6=34.18$ to 34.28 around the observed value $34.2007\pm 0.0067$ upon shifting $P_{23}$ from 5 Mbar deeper down to 18 Mbar. Taken at face value, Jupiter's observed $J_6$ value indicates that the He-depleted/He-enriched transition occurs at around 2--2.5 Mbar.

\subsection{Z-profiles for Jupiter} \label{sec:Jup_rho}

In Figure \ref{fig:zprofilesJ} we show the radial heavy element distribution of some of the models with Gaussian-$Z$ in layer No.~3. Models with unmodified H/He-adiabat appear in light-blue and are described in Section \ref{sec:Zprofiles_wo}, while models with modified H/He-adiabat appear in red and are described in Section \ref{sec:Zprofiles_w}.

\subsubsection{Unmodified H/He-adiabat}\label{sec:Zprofiles_wo}

All our models with CMS-19 EOS that fit $J_2$ and $J_4$ have negative $Z_1$ values between $-0.005$ and $-0.020$ ($-0.33$ to $-1.33$ $\times$ solar). This is consistent with \citet{HM16}, who obtained $-0.6\:\ME$ of heavy elements in the molecular region for  their MH13 EOS based model DFT-MD7.15, which, with $J_4=587$, is the one that comes closest to the Juno value of $586.61$. For  a conservative estimate of their $Z_1$ value we take 1 Mbar, the entry of their Jovian adiabat into the H/He demixing boundary of Morales et al  2013, or 2 Mbar, the pressure-medium in their H/He-rain region. With a corresponding molecular envelope mass of $\sim$30$\:\ME$ and $\sim$53$\:\ME$, respectively, we obtain  $Z_1$ between $-0.011$ and $-0.02$ for model DFT-MD7.15. In contrast, \citet{Debras19} found a variety of models for non-negative atmospheric $Z$ values of $1\times Z_{\rm Gal}=0.0167$ using CMS-19 H/He EOS.  We cannot reproduce the results of \citet{Debras19} quantitatively.

\begin{figure*}
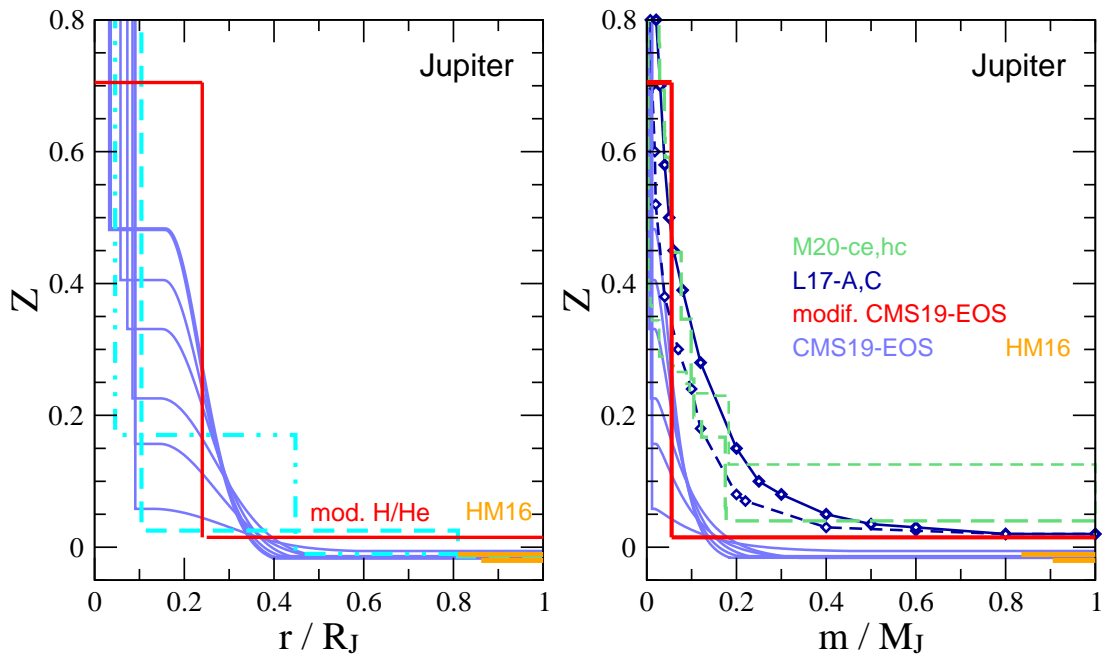

\centering
\includegraphics[width=0.40\textwidth]{f3a_zr_profilesJ.eps}
\includegraphics[width=0.40\textwidth]{f3b_zm_profilesJ.eps}
\caption{\label{fig:zprofilesJ}
Internal heavy element abundance profiles over radius (left panel) and over mass (right panel) of some of the Jupiter models in Figure 2 with Gaussian-$Z_3$ (blue), with constant-$Z_3$ (cyan) and for a model with $\rho(P)$ along the adiabat modified to yield $1\times$ solar $Z$ (red). 
Overplotted are the outer envelope Z-level of model DFT-MD7.15 from \citet{HM16} (orange), the $Z(m)$ profiles of the formation models A and C (dark blue) of \citet{Lozovsky17} at the final stage of mass accretion, after settling but before possible homogenization by mixing, and $Z(m)$ profiles of \citet{Muller20} for their  envelope accretion models assuming a hot-compact or a cold-extended state at the onset of gas accretion (green).}
\end{figure*}

The deeper the layer boundary for heavy elements is placed, the higher will the deep interior heavy element abundance become, and the smaller the core mass \citep{Nettelmann12}. With CMS-19 EOS, the response of $M_{\rm core}$ to $P_{23}$ is comparably weak, so that $P_{23}$ can be placed as deep as 20 Mbar before the compact core disappears. The thick blue model in Fig.~\ref{fig:zprofilesJ} is for $P_{23}=21$ Mbar and has a core mass of only $0.25\:\ME$. 

The non-exclusive compact core mass values of our models range from 0.2 to $4.8\:\ME$ for $P_{12}=2$ Mbar and Gaussian-$Z$ envelopes, from 0.6 to $6.0\:\ME$ for constant-Z envelopes, and from 1.2 to $3.8\:\ME$ for wind-corrected models.

For Gaussian-$Z$ deep envelopes, $Z_{3,max}$ can become quite large toward the center. We obtained $Z_{3,max}$ values up to 0.5, although larger values may be possible if the maximum of the Gaussian curve is placed at the center, while we placed it slightly off at 38 Mbar.

The mass of heavy elements in the deep interior below the negative-$Z$ envelope amounts to 7.5--10.1 $\ME$. Assuming a $1\times$ solar instead of negative-$Z$ envelope would add another $3.8\ME$ of heavy elements. A total of 11.3--13.9 $\ME$ of heavy elements is consistent with the Jupiter core accretion formation models A and C of \citet{Lozovsky17}.
These models assume solids surface densities of 6 and 10 g/cm$^2$ and planetesimal sizes of 100 km and 1 km, respectively. \citet{Lozovsky17} find that a total amount of heavy elements of $9.3\:\ME$ (A) and $16.4\:\ME$ (C) is accreted. Correspondingly, for the average $Z$-value after final mass accretion, \citet{HS17} find $\sim 0.03$--$0.05\times M_{\rm J} =$ 9.5--16$\ME$ of heavy elements for model A and a third model D, which assumes a solids surface density of 10 g/cm$^2$ like model C. In Figure \ref{fig:zprofilesJ}, models A and C are shown after settling of heavy elements but before possible convective mixing. Settling takes place if the partial pressure of ablated incoming material exceeds its vapor pressure. The resulting $Z(m)$ profiles after formation resemble our interior models with Gaussian-$Z_3$, although the Z-gradient in the post-formation models begins farther out at $\sim 0.5 M_J$ than at $\sim 0.2$--$0.3 M_J$ as in our models. On the other hand, a shallow, primordial  compositional gradient has been found to erode and to be erased in present Jupiter if vigorous convection takes place in the envelope \citep{Muller20}, while a steep compositional gradient may still persist within $0.2 M_J$. The present-state models for Jupiter of \citet{Muller20} are similar to our models with either Gaussian-$Z_3$ or constant-$Z_3$ when the heavy element-enriched deep interior (or dilute core) is assumed to begin deep inside at $> 15$ Mbar, except that our models underestimate the outer envelope metallicity while the evolution models of \citet{Muller20} overestimate it, as they yield too small a present-day radius.

\subsubsection{H/He adiabat modification}\label{sec:Zprofiles_w}

Models with negative $Z$-values are, needless to say, not considered a viable solution. There are two obvious ways how negative Z-values in the atmosphere and outer envelope can be circumvented. One possibility is to assume a superadiabatic region above the region where $J_4$ is most sensitive, which --for a polytropic Jupiter model-- is in the molecular envelope at around 50 GPa (0.9 $R_J$).  A super-adiabatic temperature profile may result from stable stratification. \citet{Christensen20} showed  that meridional flows in a stably stratified, slightly conducting region slow down the strong zonal flows and suggested the existence of such a region in Jupiter as an explanation for the truncation of the zonal flows, which, according to recent combined analysis of magnetic field and gravity field data,  occurs rather sharply at 0.97 $R_{\rm Jup}$ \citep{GalantiKaspi21}. 

However, stable stratification does not necessarily result in a super-adiabatic temperature profile. Depending on its origin, stable stratification can also be accompanied by a sub-adiabatic temperature profile. For instance, in the absence of alkali metals the opacity of the H/He fluid becomes sufficiently low for the intrinsic heat to be transported by radiation along a sub-adiabatic radiative gradient \citep{Guillot94,Guillot03}, leading to sub-adiabatic stable stratification according to the Schwarzschild criterion. Super-adiabatic gradients are predicted in a Ledoux-stable, inhomogeneous medium of upward decreasing mean molecular weight. Clouds formed by condensibles of higher molecular weight than the background composition has can induce Ledoux-stability if their abundance  is high enough. Such a scenario has been proposed for the presumably water-rich atmospheres of the ice giants \citep{Leconte17}. Water clouds may occur in Jupiter at 100 bar, silicate clouds at 1000 bar, both below the  level that so far could be probed by the Galileo entry probe (22 bar) and Juno MWR remote sensing \citep{Li20}.  Therefore, clouds are candidate causes for super-adiabatic stable stratification.

Another possibility to avoid negative metallicities is to perturb the H/He-adiabat toward lower densities. The CMS-19 hydrogen EOS shows excellent agreement with a variety of experimental data ranging from shock compression experiments for H and D at various initial conditions to isentropic compression \citep{Chabrier19}. At 50 GPa, the H EOS is even slightly stiffer than the experimental data. Only the helium EOS shows significantly higher densities in the 20 GPa to 150 GPa area than inferred from the shock compression experiments \citep{Chabrier19}.
Although the good agreement between the theoretical $P$-$\rho$ relations and the experiments as well as between CMS-19 H/He adiabats and MH13 EOS adiabats is far from suggesting that the CMS-19 H/He EOS would significantly overestimate the density along the Jupiter adiabat, we here perturb it toward lower densities. We conduct a  three-parameter study where we vary the maximum deviation $\delta\rho_{\rm max}$, the pressure entry point $P_{\rm start}$ and the pressure exit point $P_{\rm end}$. Between $P_{\rm start}$ and $P_{\rm end}$, $\delta \rho$ adopts its maximum at the logarithmic mean pressure value and is otherwise linearly interpolated as $\delta \rho(\log P)$. We explore $P_{\rm start}$ values between 1 and 50 GPa while $P_{\rm end}$ values between 50 and 150 GPa. The smaller $P_{\rm start}$ and $P_{\rm end}$, the lower the $|J_4|$ and $|J_4|/J_2$ ratio. We find $P_{\rm start} \leq 30$ GPa necessary in order to have a noticeable influence on $J_4$. Conversely, higher $P_{\rm end}$ values lead to  a stronger reduction on $J_2$. The question we ask is, for what values of $\delta\rho_{max}$, $P_{\rm start}$, and $P_{\rm end}$ can a model be found with a $1\times$ solar homogeneous $Z$?

\begin{figure*}
\centering
\includegraphics[width=0.98\textwidth]{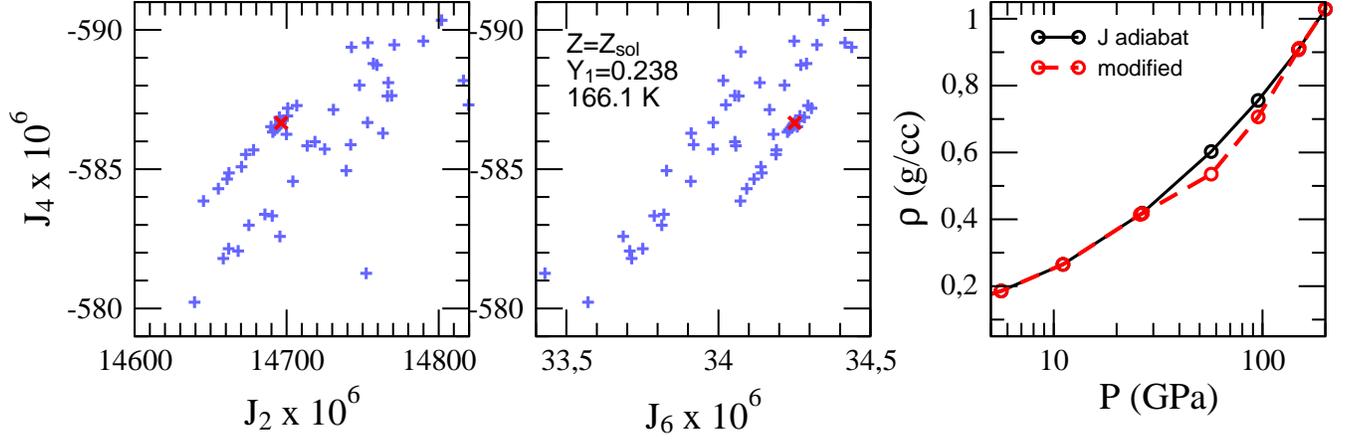}
\caption{\label{fig:perturbedJ}
$J_2$--$J_4$ (left) and $J_6$--$J_4$ (middle) values of three-layer models with $Z_1=1\times$ solar and modified Jupiter adiabats. Only the red-highlighted model meets the Juno constraints. Its adiabat is shown in the right panel (red curve).} 
\end{figure*}

For a homogeneous, unperturbed adiabat at $Y_1=0.238$ and $Y_2$ adjusted to meet $Y=0.27$, both $|J_4|$ and $J_2$ turn out significantly too large. This contrasts the result by \citet{Debras19}, who could match $J_2$ at $1\times Z_{\rm Gal}$. Thus, we need $P_{\rm start}$ to be sufficiently low for $J_4$ and $P_{\rm end}$ sufficiently high for $J_2$. We find such an optimized solution for $P_{\rm start}=26$ GPa, $P_{\rm end}=150$ GPa, and $\delta\rho=-0.1257$, i.e., a maximum reduction of the H/He adiabat by 12.57\%. The resulting $P$-$\rho$ profile is shown in Figure \ref{fig:perturbedJ}c, while the  ensemble of models in the $J_2$-$J_4$ and $J_6$-$J_4$ space is shown in Figure \ref{fig:perturbedJ}a,b.   Notably, among the wide spread of intermediate models in the $J_2$-$J_4$-$J_6$ space, the one model (red cross) that meets the Juno $J_2$, $J_4$ values, yields $J_6=34.20$, in excellent agreement with the Juno observation. For this model, the total $Z$ amounts to 15.6$\:\ME$ in good agreement with the formation model of \citet{Lozovsky17}. This exploration suggests that winds on Jupiter have a negligible influence on $J_6$.

\subsubsection{Models for enhanced 1-bar temperature}\label{sec:t1bar}

In this Section, we present Jupiter models for $T_{\rm 1bar}= 175$~K and $T_{\rm 1bar}= 180$~K, Here, we do not modify the adiabat or EOS,  and adjust the $J_2$, $J_4$ model values to the wind-corrected observed values using the corrections of \citet{Kaspi18}. 

Such warmer models are not preferred, first, because these 1-bar-temperatures significantly exceed the Galileo entry probe measurement of 166.1 K. This would not pose a problem if a mechanism had been studied that predicted a superadiabatic region underneath the 22-bar region, wherein temperatures and thus entropy would rise to the level corresponding to these or even higher 1-bar temperatures. Clouds may have a warming effect if they stabilize the region of condensation \citep{Leconte17}; however, latent heat release from condensation opposes this effect and leads to a cooler interior underneath the cloud region, as has been discussed for ice giant atmospheres \citep{Kurosaki17}. Second, jovian adiabats for different H/He EOS tend to intersect with H/He demixing curves at best in a small region at 1--3 Mbar. At present, only the rather cool MH13-EOS based Jupiter adiabat for 166.1 K shows a clear intersection by about 450 K \citep{HM16} with the state-of-the art first-principles based H/He demixing curve of \citet{Morales13}, while the intersection with the lower demixing curve $T_{\rm dmx}(P)$ of \citet{Schoettler18} is only marginal \citet{Mankovich20}. Since enhancing $T_{\rm 1 bar}$ from the Galileo value of 166.1 K by only 14 K leads to an enhancement by $\sim 350$ K at 1 Mbar and even by 460 K at 2 Mbar according to our CMS-19 EOS based Jupiter adiabats, higher surface temperatures might let the demixing region in Jupiter entirely disappear. We stress that although our CMS19 Jupiter adiabat for $T_{\rm 1 bar}=166.1$ is rather dense, it is also rather warm and with 5700 K at 1 Mbar and 6840 K at 2 Mbar outside of the first-principles based demixing regions \citep{Morales13,Schoettler18}.   

On the other hand, the recent experimentally predicted phase boundary inferred from an observed upward jump in reflectivity at 0.93 Mbar, and downward jump at 1.5 Mbar, which are interpreted as entry and exit of the compressed H/He-sample in and out from the demixing region \citep{Brygoo21} suggest high demixing temperatures of $10,000$ K. Primarily it is this finding which motivates to allow for  higher surface temperatures and for higher transition pressures, which we allow to reach the maximum where the core mass disappears, or for practical reasons drops below 1 $M_{\rm E}$.

\begin{figure*}
\centering
\includegraphics[width=0.5\textwidth]{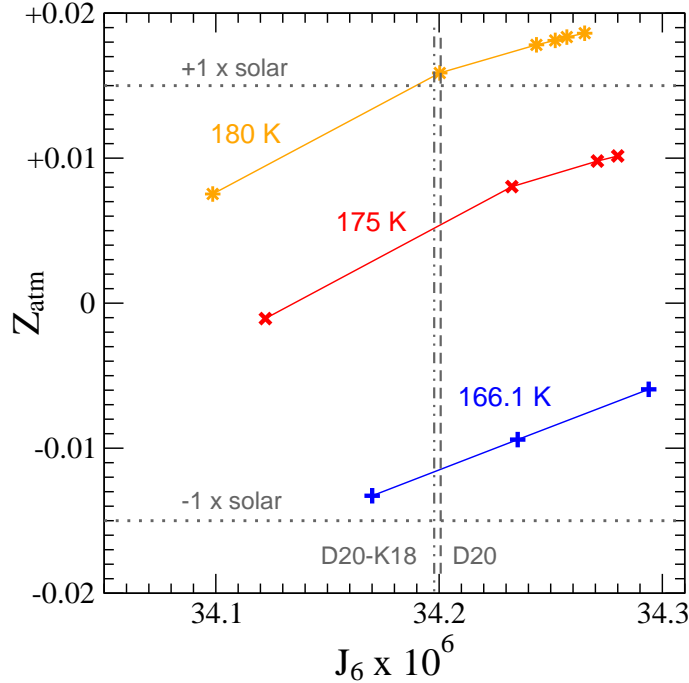}
\caption{\label{fig:ZatmJ6T1}
Outer envelope heavy element mass fraction (water) and $J_6$ values of Jupiter models with 1-bar temperature of 166.1 K for transition pressures from left to right of $P_{12}=2,3,6$ Mbars (blue), 175 K for $P_{12}$ of 2,6,8,8.5 Mbars (red), and 180 K for $P_{12}$ of 2,6,8--10 Mbars (orange). The models are calculated using the unmodified H/He-CMS19 EOS and fit the  wind-corrected $J_2$, $J_4$ values using the corrections of \citep{Kaspi18}. Horizontal dotted lines indicate $\pm 1\times$ solar metallicity $Z_{\rm solar}=0.015$, while vertical lines indicate the observed $J_6$ value of \citet{Durante20} (dashed) and its wind-corrected value (dot-dashed).}
\end{figure*}

In Figure \ref{fig:ZatmJ6T1} we show the resulting outer envelope metallicity $Z_{atm}$ and $J_6$ values. Obtaining 1x solar metallicity requires $T_{\rm 1bar}$ of 180 K (orange curve) or higher. While not negligible, additional uncertainties in the atmospheric helium abundance and $J_4$ are small and not considered here. Notably, for $T_{\rm 1 bar}=180$ K and at transition pressure between the He-poor and the He-rich region at $P_{12}=6$ Mbar, we obtain 1x solar metallicity throughout the interior down to $\sim 0.4 R_{\rm J}$, thus a largely solar-metallicity envelope. At 6 Mbar, the temperature amounts to 10,400 K at is thus at the upper limit of the experimentally inferred demixing temperature \citep{Brygoo21}. For that model, the static $J_6$ value is consistent with the observed value and its small wind-correction according to \citep{Kaspi18}. As is well known \citep{Nettelmann12}, $Z_{1}$ rises with $P_{12}$.

If there were no uncertainties in the H/He EOS, these models would suggest that the internal Jupiter adiabat lies at higher entropy than the observed adiabat down to 22 bars, and that Jupiter`s envelope metallicity is not much higher than $1\times$ solar.

\section{Application to Saturn}\label{sec:Saturn}

\subsection{Saturn models}\label{sec:Smodels}

The Saturn models of this work are built in the same manner as the Jupiter models described in Section \ref{sec:Jmodels}, although in the real planets, helium rain may induce a dichotomy \citep{Mankovich20}. We fit the Saturn models to the observed $J_2$, $J_4$ values without accounting for the wind corrections.
We assume a rotation rate of 10:32:45 hr as suggested by \citet{Helled15} would yield a best match of interior models to the observed pre-Cassini Grand Finale gravity and Pioneer and Voyager shape data. Within the given uncertainty of 46s this value is consistent with the more recently suggested rotation rates of 10:33:34~hr \citep{Militzer19}, using the Cassini Grand Finale gravity and same shape data, and with the rotation rate of 10:33:38~hr$^{+1m52s}_{-1m19s}$ inferred from the comparison of Saturn ring wave frequencies observed by Cassini with theoretical predictions for f-mode frequencies as a function of the planet's rigid-body rotation rate \citep{Mankovich19}.
We set the 1-bar surface temperature to 135 K in accordance with the Voyager measurement of $135\pm 5$~K  \citep{Lindal92}. The outer boundary is placed at a reference radius for the $J_n$ of 60330 km, which corresponds to the 0.1-bar level.

Saturn's atmospheric He abundance can be considered poorly known, as different estimates only agree in finding depletion compared to the protosolar value $Y_{\rm proto}\sim 0.27$ but disagree about the level of depletion. The lowest estimate $Y_1=0.06\pm 0.05$ stems from a combined Voyager radio occultation and infrared spectra analysis \citep{Conrath84}, while the highest estimate $Y_1=0.18$--0.25 from a reanalysis of only the Voyager infrared data \citep{CG00}. More recent Cassini data-based estimates fall in between, ranging from $Y_1=0.075$--0.13 from Cassini  infrared remote sensing \citep{Achterberg20} to $Y_1=0.158$--0.217 from Cassini stellar occultations and infrared spectra \citep{Koskinen18}. A low value of $0.07\pm 0.01$ is independently inferred from shifting the most recent H/He phase diagram of \citet{Schoettler18} to reproduce the He abundance measurement by the Galileo entry probe on Jupiter, in conjunction with the MH13 H/He-EOS and adiabat \citep{Mankovich20}. When the same procedure ia applied to the H/He phase diagram of \citet{Lorenzen11} in conjunction  with the SCvH-H/He EOS,  which both are now outdated, the yield is $Y_1=0.13$--0.16 \citep{Nettelmann15}, in between the most recent observational estimates \citep{Achterberg20,Koskinen18}. Here we construct  models for the two moderate depletion values $Y_1=0.14$ and $Y_1=0.18$ using the CMS-19 EOS and allow for a wider spread of 0.06--0.18 when using the modified CMS19-adiabat. For comparison, \citet{Galanti19} used $Y_1=0.18\pm 0.07$, \citet{Militzer19} used $Y_1=0.18$--0.26, and \citet{Ni20} used $Y_1=0.12$--0.23.  Lower $Y_1$ values yield higher atmospheric and higher maximum deep interior metallicities \citep{Militzer19,Ni20}.

Saturn thermal evolution models with H/He phase separation and helium rain predict that in Saturn, helium rains down to the core, forming a He-rich shell of 0.90-0.95 mass  percent helium atop the core \citep{Puestow16,Mankovich20}. If the helium abundance between the onset pressure of H/He phase separation and the He-shell follows a H/He phase diagram, it increases gradually with depth. However, these thermal evolution models assume for simplicity a constant low envelope metallicity. How deep He droplets can sink in a high-Z and thus higher-density deep interior, as predicted by Saturn models constrained by gravity and ring-seismology data  \citep{MaFu21}, remains to be investigated. For simplicity, we here represent the He-gradient by a jump in the He abundance at a pressure of $P_{12}=2$ to 4 Mbars deep within the He-rain region, for which predicted onset pressures in present Saturn range from 0.8 Mbar \citep{Militzer19} to 2 Mbar \citep{Mankovich20}. Below $P_{12}$, we keep the He/H ratio 
constant with depth.

Typical core-envelope pressures are around 15 Mbar. We vary $P_{23}$ between 4 and 7 Mbar, and let the Gaussian-$Z_3$ profiles adopt its maximum at 12 Mbar if the core is compact ($Z=1$) or farther out at 6--12 Mbar if the core is dilute and thus more extended. Dilute cores are created by setting  $Z=0.6$--0.7 in the core, the remaining constituent being the H/He/Z mix from envelope layer No.~3 above.

\subsection{Results for Saturn's even harmonics}

\begin{figure*}
\centering
\includegraphics[width=0.84\textwidth]{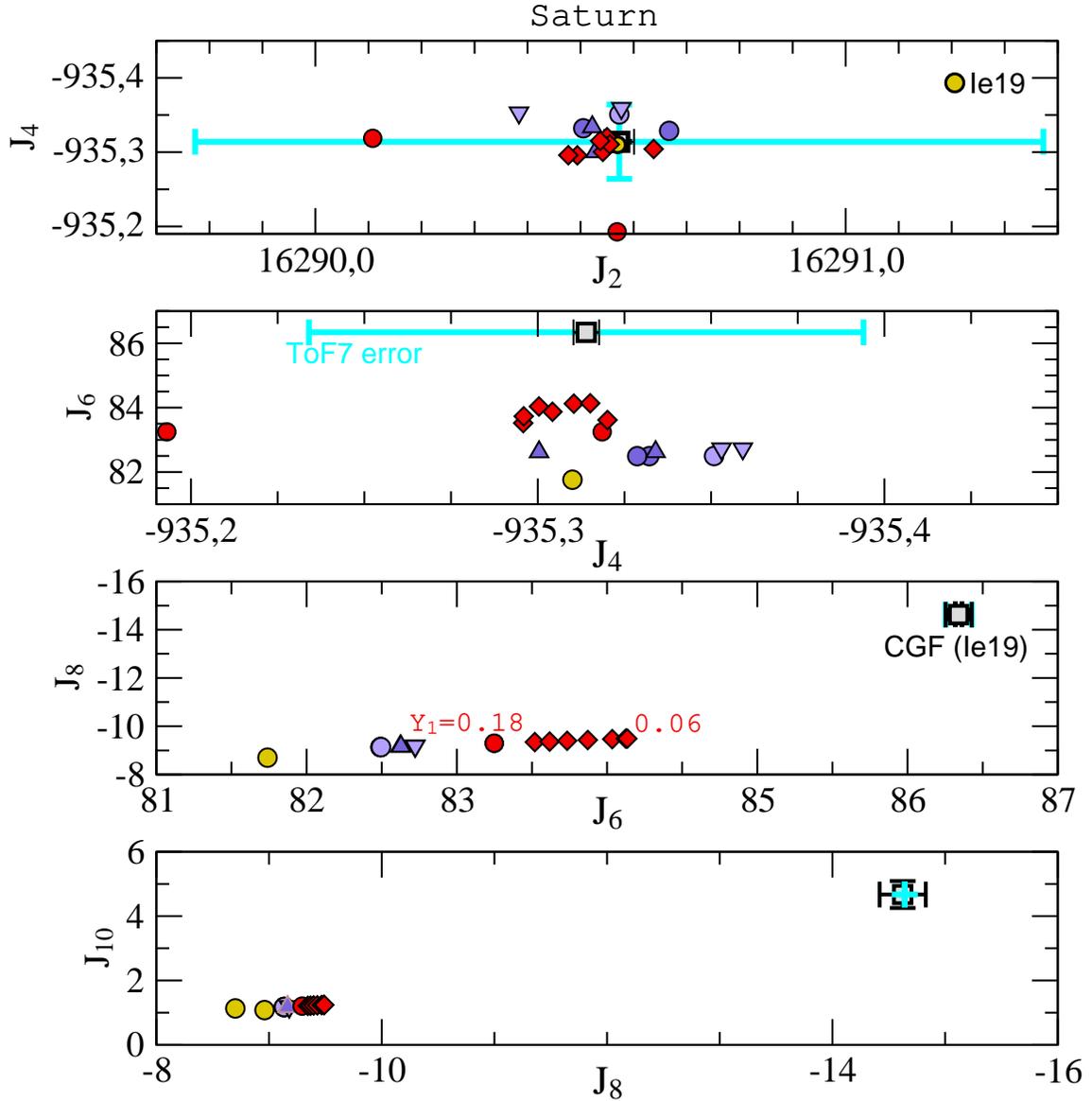}
\caption{\label{fig:satJ2n}
$J_{2n}$ values \revisedms{multiplied by $10^6$} for Saturn: Observed $J_{2n}$ values from the Cassini Grand Finale (CGR, grey squares, \citealp{Iess19}), ToF7 uncertainties overplotted to the observed values (cyan), interior model results assuming uniform rotation between 10h32m44s and 10h47m06s \citep{Iess19} (golden). This work's models: blue triangles-up: adiabatic, $Y_1=0.14$, Gaussian-$Z_3$; blue triangles-down: adiabatic, $Y_1=0.14$, constant-$Z_3$; blue circles: $Y_1=0.18$, super-adiabatic; red circles: modified H/He-adiabat, dilute core,  $Y_1=0.18$; red diamonds: modified H/He-EOS and $Y_1$ between 0.16 and 0.06,
dilute core, Gaussian-$Z_3$.
}
\end{figure*}

Figure \ref{fig:satJ2n} shows the even $J_n$ values from our Saturn models, from the two Saturn models of uniform rotation (UR) in \citet{Iess19}, and the Cassini Grand Finale observed values \citep{Iess19}.
Unlike the case of Jupiter, Saturn's observed even $|J_{n}|$ values are clearly enhanced over the model predictions for $n\geq 6$. The enhancement can be explained by rotation along cylinders that rotate approximately but not exactly with the observed speeds of clouds in the equatorial region and mid-latitude region up to $\pm 40^{\circ}$  \citep{Militzer19}. The observed $J_n$ can also be explained by a thermal wind if a little deviation of the deeper  wind speeds from the cloud speeds is allowed  \citep{Galanti19}.  The enhancement of the even $J_n$ by the zonal winds is quite substantial. Already for $J_6$, we find a 4.2--5.3\% influence, although it diminishes to 2.4--3.7\% for the modified adiabat. \citet{Iess19} obtain slightly lower UR model $J_6$ values and a 5.5\% effect, while \citet{Galanti19}, who allow for a much wider scatter in model $J_4$ values of $\pm 40\times 10^{-6 }$, obtain UR model $J_6$ values up to $87\times 10^6$, which encompasses the observed value. However, the mean of their distribution for fast, uniform rotation lies at $82\times 10^{-6}$ implying a $5.5\%$ influence of the winds on $J_6$, consistent with this work. The strong influence of the winds seen in $J_6$ suggests that also $J_4$ and $J_2$ are affected by the winds.
In Section \ref{sec:winds}, we investigate whether the observed winds are consistent with a smaller influence on $J_6$ than found in previous work \citep{Iess19}.

\subsection{$Z$- and $\rho$-profiles for Saturn}

Saturn interior models allow for higher atmospheric metallicities than Jupiter interior models when using the same H/He-EOS. For instance, \citet{Nettelmann13} obtains 1.5--6$\times$ solar for Saturn and fast rotation of 10:32:00 hr, but only 0--2.5$\times$ solar for Jupiter when applying H-REOS.2 EOS. \citet{Wahl17J} obtained only 0--$0.7\times$ solar for Jupiter using MH13 EOS, while \citet{Militzer19} obtained 1--4$\times$ solar for Saturn, consistent with \citet{Ni20} who obtained 0--$6\times$ for Saturn by considering a wide range of atmospheric He abundances, rotation rates, and wind corrections.

\begin{figure}
\centering
\includegraphics[width=0.45\textwidth]{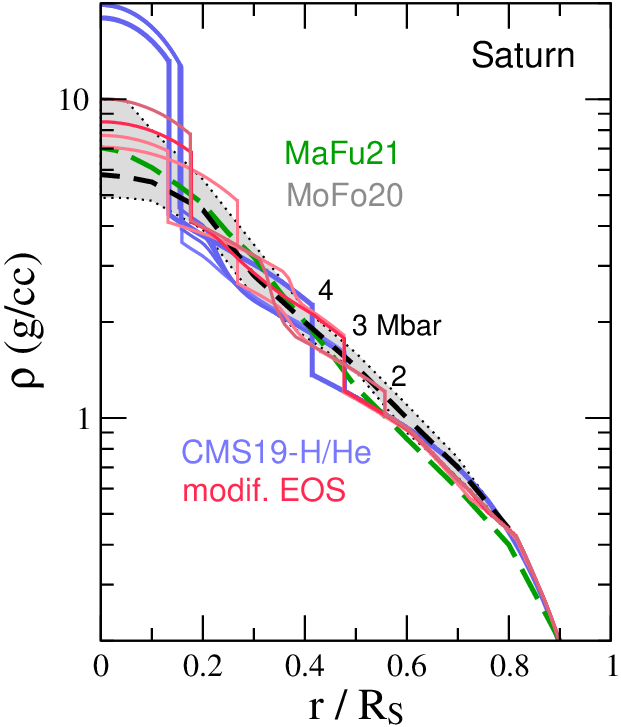}
\includegraphics[width=0.45\textwidth]{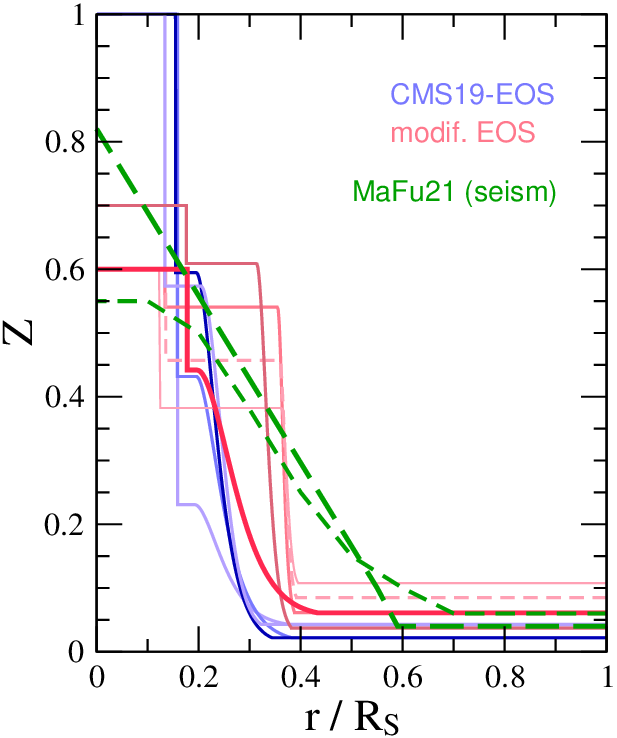}
\caption{\label{fig:sat_rho}Density profiles (left panel) and radial $Z$-profiles (right panel) of Saturn models with a compact core (blue) or a dilute core of $x_R=0.6$--0.7 and using the Jupiter-optimized adiabat (red), approximate 2$\sigma$-likelihood distribution of \citet{Movshovitz20} (grey), and  likelihood-mean from seismic constraints (green) adapted from \citet{MaFu21} assuming a linear $Z(r)$ (thick dashed) or a sigmoidal $Z(r)$ (right panel only). The highest atmospheric $Z$ levels of the red curves are for lowest atmospheric $Y=0.06$.}
\end{figure}

Here, we obtain $Z_1=0.02$--0.06 (1.5--$4\times$ solar) for nominal He-abundances $Y_1$ of 0.14--0.18, and we obtain compact non-exclusive core masses of 5--8.6 $\ME$, meaning that solutions with lower core masses are expected to be possible for deeper transition pressures than considered here. Representative $Z$- and $\rho$-profiles are shown in Figure \ref{fig:sat_rho}. Compact rocky cores yield higher central densities than suggested by the 16-84\% percentile probability range of models of \citet{Movshovitz20}, which are constrained by the gravity data. The latter models agree well with density distributions with inhomogeneous Z-profiles constrained by Cassini ring-seismology data \citep{MaFu21}. However, we were not able to find a model with a low-density core and the original H/He-EOS as such cores extend far out and yield $J_2$ values that are too large. Applying the same modification to the Saturn adiabat as for the Jupiter-optimized adiabat (see Section \ref{sec:Zprofiles_w}), we are able to obtain Saturn models with extended, low-density cores, in agreement with the likelihood distributions. As we mix H/He, with little addition of ice, into the rocky core region, we need rather high H/He amounts of 30--40\% by mass (Figure \ref{fig:sat_rho}, right panel) to reduce the core densities to 6--7~$\rm cm^{-3}$ (left panel). This is consistent with \citet{MaFu21} who need 30--40\% of H/He for rocky cores but only 0--10\% for icy cores. When also allowing the atmospheric He abundance to decrease down to 0.06, we obtain up to $7\times$ solar atmospheric $Z$. 

Even in our dilute core models, the high-$Z$ part is concentrated in the innermost 0.4 $R_{\rm Sat}$ at pressures above 4 Mbar. For comparison, seismic constraints required \citep{MaFu21} to extend the $Z$-gradient zone,  depending on the assumed functional form of $Z(r)$,  out to 0.6--0.7 $R_{\rm Sat}$, where the pressure is around 1 Mbar. While we did not explore such extended $Z$-gradients in order to keep them separated from the He-gradient zone, which we placed at at 2--4 Mbar, this comparison suggests that in the real Saturn, heavy element and helium gradients overlap. Together, the Z- and the He abundance profiles of our models suggest that the compositional gradient required to explain the ring seismology data could be due to both a diffuse core (inner region) and helium rain (outer region). 

In comparison to Jupiter, the total amount of heavy elements is clearly higher in Saturn. Models with the unperturbed H/He-adiabat yield $M_Z\sim 12.6$--$13.6\ME$ for Saturn while 7.5--$10.1\ME$ for Jupiter (Section \ref{sec:Zprofiles_wo}). Lower densities along the H/He adiabat and warmer interior temperatures would increase these values.

\section{Including the Zonal Wind profiles}\label{sec:winds}

Our Jupiter models with a modified or unmodified H/He adiabat allow for a largely homogeneous interior down to 0.4 $R_J$ with a small rock core (Figure \ref{fig:zprofilesJ}) and for $J_6$ unaffected by the winds (Figure \ref{fig:jupJ2n}). Our Saturn models allow for larger static $J_6$ values in the range (82.5--84.0)$\times 10^{-6}$ than previous work (81.8$\times 10^{-6}$), see Figure \ref{fig:satJ2n}.

Here, we investigate if such models for Jupiter and Saturn are consistent with the observed wind profiles and odd and even $J_n$ values. 
We pick one representative Jupiter model (unmodified adiabat, $P_{12}=2$, $P_{23}=20$ Mbar, compact core of mass 1.26 $\ME$) and the Saturn model highlighted in red in Figure \ref{fig:sat_rho} (modified H/He adiabat, dilute core, $Y=0.14$). 

The wind modeling approach is the same as described in \citet{GalantiKaspi21}, except that the Saturn rotation period to which the wind speeds refer is adjusted to the value of the interior model, 10h32m45s. We either use only the gravity data to constrain the wind decay depth profiles as in \citet{Kaspi18,Galanti19}, or both the gravity and magnetic field data combined (grav$+$MHD) as in \citet{GalantiKaspi21}. As an uncertainty in the static $J_{2n}$ values we take the Tof7 errors produced by \textsc{Mogrop} for $N=4000$.

\begin{figure}
\centering
\includegraphics[width=1.0\textwidth]{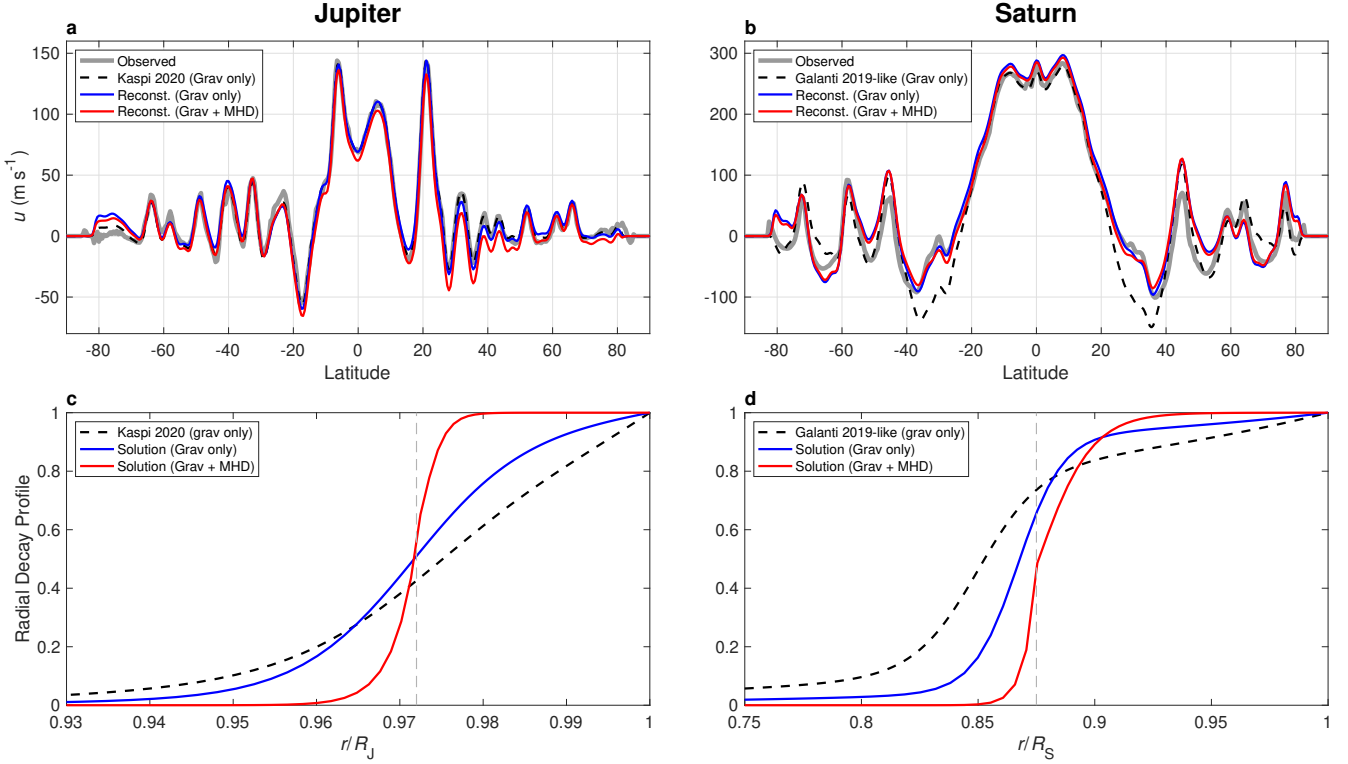}
\caption{\label{fig:winds} Wind profiles (top) for Jupiter (left) and Saturn (right) using the constraints from observed gravity data only (blue) or also from MHD (red). The black dashed lines for Jupiter are taken from \citet{Kaspi18} (grav only), while for Saturn, they are adjusted from \citet{GalantiKaspi21} (grav only). Solutions for grav+MHD from that previous work are not shown since the solutions are close to the ones from this work. The grey lines in the top panels show the observed profiles \citep{Tollefson17}.}
\end{figure}

With this approach, we are able to find fits that reproduce all the observed $J_n$ within the observational uncertainties. The wind profiles and decay depths are shown in Figure \ref{fig:winds}. This implies that extending the wind profiles, roughly as they appear at the cloud level, gives a good match to the difference between the Juno and Cassini measurements and our preferred models. Nonetheless there is enough freedom in these solutions that other wind profiles with small shifts to the wind profiles can give fitting solutions as well \citep{Galanti21}. For these wind profiles, grav and grav+MHD yield similar solutions. Jupiter's wind profile is slightly less well matched (red and blue lines more strongly deviate from the observed profile (grey) than does the black-dashed line), while for Saturn, the shoulders at 20-40 degrees latitude are somewhat better matched than in previous work \citep{Iess19,Galanti19}. The wind decay depths for grav only are slightly steeper than for previous interior models and thus closer to the grav+MHD solutions.

For each of Jupiter and Saturn, we picked only one specific interior model to calculate the wind contribution and optimize for the agreement with the observed wind velocities and gravitational harmonics. The fact that these two interior models allowed for solutions within the observed values and the ToF7/\textsc{Mogrop} uncertainty strongly suggests that there are further interior models for which such a fit can be obtained. This means that the joint interior and wind solutions are not unique, given the uncertainties we allowed for. In addition, alternative interior models which fit all the $J_n$ when combined with a wind model may be possible for different equations of state and wind models, such as the MH13 H/He EOS and wind models that account for the oblate shape \citep{Cao17} or solve for the gravo-thermal wind equation \revisedms{that accounts for the dynamic self-gravity of the flow (TGWE)} \citep{Kong18,Wicht20}. \revisedms{We note that \citet{GalantiTziperman17} find that these modified wind models introduce corrections that are an order of magnitude smaller for most $J_n$, while \citet{Dietrich21} obtain corrections of respectively $\sim$60\% and $\sim$20\% for $J_3$ and $J_5$ when including the dynamic self-gravity for polytropic models and an additional correction of $\sim$40\% and $\sim$10\% when accounting for Jupiter-model specific background density and gravity profiles}.

\revisedms{
Internal flow structures, that are decoupled from the observed cloud-level winds, can also be found to fit the $J_n$ \citep{Kaspi18, Kong18}} and thus lead to non-uniqueness of the solutions \citep{Kong18}. Here, we conclude for non-uniqueness because of uncertainties in the interior models, the high-order $J_n$ to be fitted, and the wind profile.

\section{Discussion}\label{sec:discus}

\revisedms{
Our Jupiter and Saturn models exhibit a strong trend toward low envelope metallicities which extend deep into Saturn's interior to $\sim 0.4\: R_{\rm S}$ (Fig.~\ref{fig:sat_rho}) or are negative in  Jupiter (Fig.~\ref{fig:zprofilesJ}). We have attributed these model properties to possible uncertainties in the H/He EOS, however, one may think of further processes.
}

\subsection{$Z$ and the adiabatic $P$--$T$ profile}

\revisedms{
We did not include $Z$ in the computation of the adiabatic $P$-$T$ profile. This leads to a slight  overestimation of the temperatures along the adiabat. Mixing first the equations of state H/He-REOS and H2O-REOS linearly and then computing the adiabats as a function of $Z$ using thermodynamic integration described in \citet{Nettelmann12} shows that 10x solar water would lower the temperatures by only $-100$ K in the 10-100 GPa region relevant for $J_2$ and $J_4$. Conversely, an adiabat more rich in atomic helium would be warmer. Considering molecular volatiles in the entropy calculation would tend to make the adiabat slightly cooler and denser, and therefore lead to even lower envelope metallicities, but our estimate shows this effect should be small. 
}

\subsection{H/He demixing?}
\revisedms{
Inspired by the recent experimentally derived H/He demixing boundary that extends
over a large region from $\sim$0.9 Mbar to $\sim$10000 K (8--10 Mbar) in Jupiter \citep{Brygoo21} one could consider helium abundances which increase over a wide region, allowing for more heavy elements to replace helium. However, our variation of the H/He adiabat showed that reduced densities are needed near the top and beyond ($\sim$20 GPa) the demixing region. Exploration of the helium abundance profile deep inside may thus have too little influence to solve the low-atmospheric metallicity problem in Jupiter.
}

\subsection{Deep internal flows in Jupiter?}

\revisedms{
\citet{Guillot18} constrained the maximum amplitude of a deep wind that would extend along cylinders all the way to the center and be consistent with the even $J_{n}$ to $<10$ m/s. \citet{Kong18} found that a flow with 1 m/s down to $0.8\: R_{\rm J}$ can explain the odd $J_n$; but including the influence of the induced magnetic field through Ohmic heat dissipation bounded by the total convective power \citep{Wicht19}, \citet{LiKong20} are able to limit this depth to only $0.96 R_{\rm Jup}$ results. \citet{Moore19} even constrain the flow velocity to a few mm/s at depths of 0.93--0.95 $R_{\rm J}$ by explaining the observed secular variation of the magnetic field with advection by the flow.}\\
\revisedms{
In contrast, in order to lift Jupiter's atmospheric metallicities substantially, a much stronger and retrograde deep wind in the interior where $J_2$ and $J_4$ are sensitive would be needed. This deep wind must not be seen in the high-order gravity data, in the secular variation of the magnetic field, nor in the System III rotation period derived from magnetic field observations. It would be seen in the moment of inertia, the static Love numbers, and the shape. At present, there is no indication of a strong ($> 10$ m/s) wind in Jupiter's deep interior.}

\subsection{Uncertainty in the shape due to dynamical effects?}

\revisedms{
With both CMS and ToF-method the interior models are derived from a self-consistent, static solution between the gravity field and the shape, however, the shape and the gravity field of the planet can be influenced by various dynamic effects.}\\

\revisedms{ 
For instance, \citet{KZ2020} propose that the winds are shallow while convective motions could induce a zonal flow disjunct from the surface winds. They find a dynamic influence of 1x$10^6$ in $J_2$ and 0.2x$10^6$ for $J_4$. While small, this effect on $J_4$ could be noticeable in the interior models. However, this estimate of the dynamic contribution due to convective motions is based on an Ekman number 5x$10^{-5}$, about 10 orders of magnitude larger than in the real Jupiter and Saturn. It is therefore possible, that the dynamic contribution from convective motions on the low-order $J_{2n}$ is smaller in the real planets.
} \\

\revisedms{
For Saturn, the uncertainty in its deep rotation rate maps on an uncertainty in equipotential shape of about 120 km \citep{HellGui13}, far outweighting other influences like from the winds, which lift the dynamical height above a reference iso-bar to no more than $\sim 20$ km \citep{Buccino20}.
Moreover, the zonal flows on  Saturn are symmetric enough to be described by rotation along cylinders up to mid-latitudes \citep{Militzer19}. In that case, equipotential theory still applies. We do not suggest that dynamic effects play a major role for the uncertainty in Saturn`s shape and gravity field.
} \\

\revisedms{
For Jupiter, the uncertainty in rotation rate is tiny, so that it is the influence of the winds of 2--4 km \citep{Buccino20} against which further effects must be compared. Such are the tidal buldges from the Galilean satellites.  \citet{Nettelmann19} estimated a maximum elongation of 28 km in the direction of Io from static tidal response.
The tidal flows around Jupiter are a dynamic perturbation and subject to Coriolis force \citep{Idini21, Lai21}. The flow and the Coriolis force acting upon it lead to dynamic contributions to the Love numbers $k_{nm}$.  JUNO measurements revealed a deviation by 1--7\% from the static $k_2$-value \citep{Idini21}. Approximating the corresponding shape deformation $h_2$ by $h_2=1+k_2$ yields a tentative estimate of a (1--7)\% $\times 28$ km $\sim$(3--21) km additional shape deformation due to dynamic tidal response, which exceeds the wind effect. Possible importance of (periodic) perturbations on Jupiter`s  interior structure inference remains to be investigated. 
}

\subsection{A cold hot spot?}
\revisedms{
Juno MWR data revealed that the ammonia abundance below the cloud level shows strong vertical and latitudinal variation \citep{GuillotFletcher20}. It is therefore possible that Jupiter's atmosphere is not everywhere well mixed where observations were taken. Consequently, the abundances and temperatures measured by the Galileo entry probe in a hot spot may not be representative of Jupiter's global atmosphere. On the other hand, analysis of Voyager 1 and 2 radio occultation data spanning a broad range of latitudes between 70 degrees South and the equator, yielded a 1-bar temperature of 165 K +/- 5 K \citep{Lindal81}, consistent with the Galileo measurement of 166.1 K in the hot spot. Present data therefore do not indicate that hot spots, in which deeper layers are exposed that appear brighter than surrounding regions at higher altitudes, were particularly cool regions allowing us to suppose warmer global average temperatures. 
Rather, it is possible that the hot-spot temperature-gradient is steeper than the global one since it is close to a dry adiabat \citep{Seiff1998}, whereas moist regions above the water cloud level may follow a less steep $P$--$T$ profile \citep{Kurosaki17}, implying an even cooler interior below the cloud base. A colder and thus denser interior would strain the low-metallicity models even more. 
}

\section{Conclusions}\label{sec:conclusions}

We present the expansion of the Theory of Figures \citep{ZT78} from formerly 5th order \citep{ZT75} to the 7th order. The coefficients are available in form of five read-in online tables and allow the computation of the even gravitational harmonics $J_2$--$J_{14}$ and the shape of a rotating fluid body in hydrostatic equilibrium.

We  estimate the numerical accuracy of the ToF method carried out to 4th \citep{Nettelmann17},  5th \citep{ZT75}, and 7th (this work) order by comparing to the analytic Bessel solution of \citet{WH16} for the rotating $n=1$ polytrope and by using three different codes. We find that the CEPAM code \citep{GuillotMorel95} with ToF5 \citep{Ni20} has a superior performance in regard to the accuracy in $J_2$, $J_4$, and $J_6$, while for $J_8$ and $J_{10}$, the \textsc{Mogrop} code with ToF7 reaches similar degree of accuracy for a practical number of radial grid points of a few thousand, although in $J_{10}$ the error changes sign between both variants.
The accuracy in $J_8$, $J_{10}$, $J_{12}$ falls by, respectively, 1, 2, 3 orders of magnitude below the current 3$\times$ $1\sigma$ formal uncertainty of the observational gravity data ''halfway through the Juno  mission'' \citep{Durante20}.
We also apply the CMS-2019 H/He EoS of \citet{Chabrier19} to interior models of Jupiter and Saturn. 
\par 
For Jupiter, the high-order $J_n$ of the Jupiter models fall along the same line in $J_n$--$J_{n+2}$ space as in previous work, regardless of detailed model assumptions and the H/He EOS used. We find that $J_6$ stands out in that it is neither adjusted, as $J_2$ and $J_4$ are,  nor insensitive to model assumptions, as the $J_n$ for $n\geq 8$ are. We match Jupiter's observed $J_6$ value by placing the transition pressure between an outer, He-depleted envelope and an inner, He-enriched envelope at $P_{12}=2$--2.5 Mbar. Transition pressures farther out lead to lower $J_6$ values, while deeper transitions result in higher $J_6$ values. The same behavior but with a weaker amplitude is seen for the transition pressure of heavy elements, which we place between $\sim 5$ and 20 Mbar.  
Gaussian-$Z$ profiles underneath can lead to high metallicities of up to Z=0.5 at the compact core-mantle boundary. However, the atmospheric heavy element abundance, represented by an EOS of water, always stays negative ($\sim -1\times$ solar) if the adiabat is defined by the 1-bar temperature of 166.1 K as measured by Galileo and extended downward.  Alternatively, we set $Z_1$ to $1\times$ solar, the 1$\sigma$-lower limit of the equatorial water abundance measured by Juno \citep{Li20} and perturb the adiabat to fit $J_2$ and $J_4$. Such an optimized adiabat was found for a perturbation between 26 and 150 GPa and has a maximum density decrease of 12.6\% at a midpoint of 63 GPa. Higher internal temperatures help to decrease the internal density as well. For $T_{1 bar}=180$ K and deep transition pressure $P_{\rm trans, He}=6$ Mbar we obtain 1x solar metallicity without H/He-EOS modification.

Our Saturn models with CMS-19 H/He EOS are characterized by a few-times solar envelope that extends deep down to $< 0.4\: R_{\rm Sat}$ and requires a compact core. Its density of $\sim 20$ g$\rm cm^{-3}$ is higher than the most likely central densities of Saturn that match the gravity field \citep{Movshovitz20}, which in turn agree with density distributions of a largely stably stratified deep interior with a dilute core \citep{MaFu21}.  By applying the Jupiter-optimized perturbation along the adiabat to Saturn, we are able to obtain density distributions with a dilute core of 30-40\% H/He that reaches out to $\sim 0.4 R_{\rm Sat}$ in the core. This moves the solution in the direction of density distribution constrained by seismic data. Our models suggest that an inhomogeneous central region out to $\sim 0.6$ $R_{Sat}$ \citep{MaFu21} is due to both a dilute core and rained down helium.

\revisedms{
Overall, our Jupiter and Saturn models exhibit a strong trend toward low envelope metallicities which extend deep into Saturn's interior to $\sim 0.4\: R_{\rm S}$ (Fig.~\ref{fig:sat_rho}) or are negative in  Jupiter (Fig.~\ref{fig:zprofilesJ}). We have attributed these model properties to possible uncertainties in the H/He EOS. However, further processes one may think of and which certainly are at play are estimated to be too minor to solve that issue.
}

This work demonstrates that our understanding of the internal heavy-element distribution of Jupiter and Saturn strongly depends on the properties of H and He. We conclude that part of the difficulties of obtaining Jupiter and Saturn models that are consistent with all observational constraints still lies in our imperfect understanding of the material properties. We therefore suggest that further measurements and calculations of the behavior of materials at planetary conditions could improve our understanding of the gas giants. This, in return, will also reflect on the characterization of gaseous planets orbiting other stars.

\begin{acknowledgments}
We thank the IWG members of the JUNO Team for discussions.
NN and JJF acknowledge support through NASA's Juno Participating Scientist Program under grant 80NSSC19K1286. \revisedms{We thank the two anonymous reviewers for the constructive reports and insightful comments.}
\end{acknowledgments}

\appendix

\section{Technical Notes on ToF coefficient computations}
\label{sec:app_ToFnotes}

In the following, we abbreviate the term in brackets in Eq.~(\ref{eq:rl}) as $(1+\Sigma)$ and write $\mu=\cos\vartheta$. Since any point $\mathbf{r}=(r,\varphi,\vartheta)$ in and near the planet can be associated with an equipotential surface, we can replace any dependence  $f(r,\vartheta)$ by $f(l,\vartheta)$ 
using Eq.~(\ref{eq:rl}).

\subsection{Centrifugal Potential $Q$}

The first step of the Legendre polynomial expansion of the centrifugal potential 
$Q=1/2\,\omega^2\,r^2\sin^2\vartheta$ is to write $Q=-1/3\,\omega^2\,r^2 (P_0 - P_2(\mu))$. Its full expansion is obtained by replacing $r$ with $r_l(\vartheta)$. $Q(l,\vartheta)$ can then be written in the form
\begin{equation}
	Q(l,\vartheta) = -\,\frac{GM}{R_m}\:\left(\frac{l}{R_m}\right)^2\sum_{k=0}^{O} A_{2k}^{(Q)}P_{2k}(\mu)\quad
	\mbox{with}\quad A_{k}^{(Q)}=\frac{m_{\rm rot}}{3}\:\sum_{i}c_{i\,0\,k}\:.
\end{equation}

\subsection{Gravitational Potential $V$}

The gravitational potential $V(\mathbf{r})=-G\int d^3r'\:\rho(\mathbf{r'})/|\mathbf{r'}-\mathbf{r}|$ 
can be separated into an external contribution $D$ from the mass density interior to a sphere of radius $r$,  i.e.~to which $r$ is exterior ($r>r'$), and an internal contribution $D'$ from the mass density exterior to a sphere of radius $r$, i.e.~to which $r$ is interior ($r<r'$). $V$ then reads
\begin{equation}\label{eq:Vr}
	V(r,\vartheta) = -G \sum_{n=0}^{\infty}
	\left( r^{-(n+1)}D_{2n}(r) + r^n D_{2n}'(r) \right)\:P_{2n}(\mu)\quad
\end{equation}
with 
\begin{equation}\label{eq:Dn}
	D_n = \int_{r'<r} d^3r' \rho\: \:r'^{\,n}\:P_n(\mu')\quad, 
	D_n' = \int_{r'>r} d^3r' \rho\: \:r'^{\,-(n+1)}\:P_n(\mu')\:. 
\end{equation}
Although this multipole expansion is valid only for spheres, in ToF, the radial coordinate $r$ in Eqs.~(\ref{eq:Vr},\ref{eq:Dn}) is simply replaced by the non-spherical equipotential surface $r_l(\vartheta)$
and the interior/exterior criterion is transferred to $l$. In the CMS method, this expansion is also used but the expression for the external potential is only applied to spheroids of level surface $r_i(\mu)$ at or interior ($i\geq j$) to a point B of radial distance $r_B$ that resides on a level surface $r_j(\mu)$.  Since all spheroids share the same center but extend outward to different level surfaces $r_i(\mu)$ 
where $i=0$ denotes the surface of the planet and $i=N$ the center, a point B on $r_j(\mu)$ is also  located exterior to the mass of the spheroids of index $i<j$ but only as far as the radius $r_B$. 
This is taken care of in the CMS method by adding the external gravitational potential of the spheres of densities $\delta_{i,\, i<j}$ interior to $B$ from the spheroids $i<j$. This improvement of the CMS method over ToF method 
is still limited by the deformation and spacing of the spheroids. Rapid rotation, or dense spacing, could lead to an overlap of the sphere of radius $r_j$ with the spheroid $r_{j-1}(\mu)$. 
\citet{Kong13} developed the full solution to the Poisson equation and demonstrated that the CMS method  converges as long as the flattening ($R_{eq}/R_{pol} -1$) remains sufficiently small. Similarly, 
\citet{Hubbard14} showed that ToF converges for sufficiently small flattening and toward the correct solution.

By replacing all powers of $r$ by $r_l(\vartheta)$ in Eq.~(\ref{eq:Vr}), one obtains
\begin{equation}\label{eq:Vl}
	V(l,\vartheta) = -\frac{GM}{R_m}\left(\frac{l}{R_m}\right)^2 \sum_{n=0}^{O}
	\left( (1+\Sigma)^{-(n+1)}S_{2n}(z) + (1+\Sigma)^n S_{2n}'(z) \right)\:P_{2n}(\mu)\quad
\end{equation}
with $z=l/R_m$, $z\:\epsilon\: [0,1]$ and, with the help of the transformation 
\[
	r^n\:dr = dl\: r^n \frac{dr}{dl} = \frac{1}{n+1} \: dl\: \frac{d}{dl}\: r^{n+1}\quad,
\]
\begin{eqnarray}\label{eq:Sn}
	S_n(z) &=& \frac{3}{2(n+3)}\:\frac{1}{z^{n+3}} \int_0^z dz'\: \frac{\rho(z')}{\bar{\rho}}\: 
	\frac{d}{dz'}\left[\:z'^{\,n+3} \int_{-1}^1 d\mu'\:(1+\Sigma)^{n+3}\:P_n(\mu')\right]\quad, \\
	S_n'(z) &=& \frac{3}{2(2-n)}\:\frac{1}{z^{2-n}} \int_z^1 dz'\: \frac{\rho(z')}{\bar{\rho}}\:
	\frac{d}{dz'}\left[z'^{\,2-n} \: 	\int_{-1}^1 d\mu'\: (1+\Sigma)^{2-n}\:P_n(\mu')\right]\quad.
\end{eqnarray}
This can further be written as
\begin{eqnarray}\label{eq:Snfn}
	S_n(z) &=& \frac{1}{z^{n+3}} \int_0^z dz'\: \frac{\rho(z')}{\bar{\rho}}\:
	\frac{d}{dz'}\left[\:z'^{\,n+3}\:f_n(z)\right]\quad, \\
\label{eq:Snpfnp}
	S_{n,}'(z) &=& \frac{1}{z^{2-n}} \int_z^1 dz'\: \frac{\rho(z')}{\bar{\rho}}\:
	\frac{d}{dz'}\left[ z'^{\,2-n}\:f_n'(z)\right] \quad,
\end{eqnarray}
with 
\begin{eqnarray}\label{eq:fn}
	f_n(z) &=& \frac{3}{2(n+3)}\int_{-1}^1\! d\mu\: P_n(\mu)\:(1+\Sigma)^{n+3}\:,\nonumber\\
	f'_n(z) &=& \frac{3}{2(2-n)}\int_{-1}^1\! d\mu\: P_n(\mu)\:(1+\Sigma)^{2-n} \quad (n\not=2)\,,\nonumber\\
	f'_2(z) &=& \frac{3}{2}\int_{-1}^1\!d\mu\: P_n(\mu)\:\ln(1+\Sigma)\quad.
\end{eqnarray}

\subsection{ToF7 Tables for public usage}\label{sec:app_tables}

The ToF coefficients $c_{ink}$ are of the form 
\[
	c_{ink} = q_{ink}\: \prod_{j=1}^{O} s_{2j}^{p_{2j,\,ink}}\:,
\]
where the $q_{ink}$ are rational numbers and the exponents $p_j$ are small natural numbers including 0.

Since the number of coefficients rises with the order of expansion faster than quadratically, it becomes impractical to write down all the coefficients. We present them in the form of five online tables. Tables {\tt tab\underline{~}Sn} and {\tt tab\underline{~}Snp} contain the coefficients $c_{ink}$ and $c_{ink}'$ in front of the $S_n$ and $S_n'$ in the $A_k^{(V)}$, respectively, so that 
\[
	A_k^{(V)} = \sum_{i=1}^{N_{0\,k}} c_{i\,0\,k} S_0 + \sum_{i=1}^{N_{2\,k}} c_{i\,2\,k} S_2 +\ldots 
	+  \sum_{i=1}^{N_{14\,k}} c_{i\,14\,k} S_{14} 
	+ \ \sum_{i=1}^{N_{0\,k}'}  c_{i\,0\,k}' S_0' + \sum_{i=1}^{N_{2\,k}'}  c_{i\,2\,k}' S_2' +\ldots 
	+  \sum_{i=1}^{N_{14\,k}'}  c_{i\,14\,k,}' S_{14}' 
\]
Table {\tt tab\underline{~}m} contains the summands in the $A_k^{(Q)}$ so that $A_k^{(Q)} = m_{\rm rot}/3\:\sum_{i=1} c_{i\,0\,k}$ with $m_{\rm rot}=\omega^2\,R_{m}^3/GM$.
Tables {\tt tab\underline{~}fn} and {\tt tab\underline{~}fnp} contain the summands of the functions $f_n$ and $f_n'$ so that $f_n = \sum_i c_{in}$, $f_n' = \sum_i c_{i\,n\,}'$, respectively.
In Table \ref{tab:Sn}, we give an example of the read-in ascii-table {\tt tab\underline{~}Sn}.
All five tables have the same format. 
\begin{table}
\centering
\begin{tabular}{lll llllll | c}
$n$  &  $k$ & $N_{nk}$  &&&&&&& \\ 
 Order  &  $p_2$ & $p_4$ &$p_6$ &$p_8$ &$p_{10}$ &$p_{12}$ &$p_{14}$ & $q_{ink}$, $i=1$---$N_{nk}$ & comment\\
\hline
 0  &0  &24 &&&&&&& $n=0$, $k=0$, next 24 rows\\
 0    &0  &0  &0  &0  &0  &0  &0    &1.000000000000000e+00  & Order$=0$, $c_{1\,0\,0}=1$\\
 2    &2  &0  &0  &0  &0  &0  &0    &4.000000000000000e-01  & Order$=2$, $c_{2\,0\,0}=0.4\, s_2^2$\\ 
 \vdots &&&&&&&&\\
 7    &2  &1  &1  &0  &0  &0  &0    &2.157842157842158e-01 & Order$=7$, $c_{24\,0\,0}=q_{24\,0\,0}\: s_2^2s_4^1s_6^1$\\
 \multicolumn{9}{c}{\ldots} &\\
 4  &8 &17 &&&&&&& $n=4$, $k=8$, next 17 rows  \\
 4    &2  &0  &0  &0  &0  &0  &0    &2.937062937062937e+00 & Order$=4$, $c_{1\,4\,8}=q_{1\,4\,8}\: s_2^2$\\
 \vdots &&&&&&&&
\end{tabular}
\caption{\label{tab:Sn} 
Example of one of the five online-only ascii-tables. This example is for Table {\tt tab\underline{~}Sn} that contains the coefficients in front of the $S_n$ in the $A_k$. 
Since the $A_{2k}^{(Q)}$ have no dependence on index $n$, index $n$ is set to 0 in Table {\tt tab\underline{~}m}. Since the $f_n$, $f_n'$ have no dependence on index $k$, $k$ is set to 0 in Tables {\tt tab\underline{~}fn} and {\tt tab\underline{~}fn'}.}
\end{table} 
The $A_{2k}$ in Eq,~(\ref{eq:U_Ak}) are of the form $0 = A_{2k}(l) = - s_{2k}(l)S_0(l) + B$ if $k\neq0$. We rewrite this as an expression for direct, iterative computation of the figure functions in the form $s_{2k}=B/S_0$, where the functions $B$ depend on the $\{s_{2k}\}$ from the previous iteration step. We  omit the summand $-s_{2k}S_0$ from Table {\tt tab\underline{~}{Sn}}.

\revisedms{To facilitate the application of our ToF7 tables by external users for their own planetary models, we share routines for read-in of the tables in Matlab, Python, and C$++$. For the latter variant, we also provide functions that can be used to easily access the coefficient values. The archived service routines and descriptions can be found at https://doi.org/10.6084/m9.figshare.16822252. }

\subsection{\revisedms{Powers of the radius}}

In the binomial expansion of $(1+x)^{-m}$ for $m>0$,
\begin{equation}  
	(1+x)^{-m} = \sum_{i=0}^{\infty}\left(\!\!\!\begin{array}{c}-m\\ i\end{array}\!\!\!\right)x^i\quad,
\end{equation}
it is sufficient to expand to $i=7$ because $x=\Sigma$ and the minimum order of $\Sigma^i$ is $i$. 
The binomial expansion of $(1+x)^{m}$ for $m>0$,
\begin{equation}  
	(1+x)^{m} = \sum_{i=0}^{m}\left(\!\!\!\begin{array}{c}m\\ i\end{array}\!\!\!\right)x^i\quad,
\end{equation}
is carried out to $m\leq 7$. Products $P_n P_m$ occurring in $\Sigma^i$ and in $(1+\Sigma)^i P_j$ are expanded as $\sum_{k=0}^{n+m} b_kP_k$. It becomes evident that all terms can linearly be expanded in 
Legendre polynomials and that numbers in the expansion coefficients are rational numbers $q = n_{\rm e}/n_{\rm d}$.
The natural numbers  $n_{\rm e}$, $n_{\rm d}$ can be represented exactly on a computer, although size limitations may apply.  For ToF7, the vast majority of numbers could be decomposed into prime numbers that
individually do not exceed 3 million. However in rare cases this was not possible and larger prime numbers would have been required, perhaps indicating an error in the code used. In such a case, the given number is not decomposed into prime numbers. In any case, the enumerators and denominators are computed as exact numbers in all coefficients. They are cast to real numbers of 15 digits only for the purpose of printing 
the tables.

\subsection{Figure function $s_0$}

We calculate $s_0$ to the 7th order with the help of the defining integral Eq.~(\ref{eq:equalvoll}),
which, with $z=l/l_1$, can be written
\begin{equation}\label{eq:s0integral}
	\frac{4\pi}{3}1_1^3 = 2\pi\,l_1^3\int_{-1}^{1}d\mu\:\int_0^1 dz\: z^2(1+\Sigma(z,\mu))^3\quad.
\end{equation}
Because of hemispheric symmetry and $1/3=\int_0^1 dz\: z^2$, the comparison of integrands yields 
\begin{equation}\label{eq:s0integral1}
	1 = \int_{0}^1  d\mu\: (1+\Sigma)^3\quad,
\end{equation}
where $\Sigma(z,\mu)=\sum_0^7 s_{2i}(z)P_{2i}(\mu)$ and $(1+\Sigma)^3 = 1 \:+\: 3\Sigma \:+\: 3\Sigma^2 + \Sigma^3$.
We calculate $\Sigma^3$ and safely remove all terms containing $P_0P_nP_m$ with $n \not{=}m$ or $P_0^2P_n$,
since they would contribute nothing to the evaluated integral in Eq.~(\ref{eq:s0integral1}),
\begin{eqnarray}
	\Sigma^3 &=& s_0^3P_0^3 \:+\: s_2^3P_2^3 \:+\: s_4^3P_4^3 \nonumber\\
	&& +\: 3s_0s_2^2P_0P_2^2 \:+\: 3s_2^2s_4P_2^2P_4 \:+\: 3s_2^2s_6P_2^2P_6 \:+\: 3s_2^2s_8P_2^2P_8 
	\:+\: 3s_2^2s_{10}P_2^2P_{10} \nonumber\\
	&& +\: 3s_0s_4^2P_0P_4^2 \:+\: 3s_2s_4^2P_2P_4^2 \:+\: 3s_4^2s_6P_4^2P_6
	+\: 3s_2s_6^2P_2P_6^2  \nonumber\\
	&& +\: 6s_2s_4s_6P_2P_4P_6  \:+\: 6s_2s_4s_8P_2P_4P_8  \quad.
\end{eqnarray}
Terms containing $P_0P_n^2$ yield a factor $1/(2n+1)$ for the integral in (\ref{eq:s0integral1}). By $(P_nP_m)_k$ we denote the summand $b_kP_k$ in the expansion of $P_n\times P_m$. The other terms contribute
\begin{equation}\begin{array}{lll}\label{eq:sigma3_7th}
P_0^3 &: & \to 1\\
P_2^3 &: (P_2P_2)_2P_2 = \frac{2}{7}P_2^2 &\to \frac{2}{5\cdot 7}\\
P_4^3 &: (P_4P_4)_4P_4 = \frac{2\cdot 3^4}{7\cdot 11\cdot 13}P_4^2 &\to \frac{2\cdot 3^2}{7\cdot 11\cdot 13}\\
P_2^2P_4 &: (P_2P_2)_4\:P_4 = \frac{18}{35}P_4^2 & \to \frac{2}{5\cdot 7}\\
P_2^2P_6 &: (P_2P_2)_6\:P_6 & \to 0\\
P_2^2P_8 &: (P_2P_2)_8\:P_8 & \to 0\\
P_2^2P_{10} &: (P_2P_2)_{10}\:P_{10} & \to 0\\
P_2P_4^2 &: P_2(P_4P_4)_2 = \frac{2^2\cdot 5^2}{7\cdot 11}P_2^2 & \to  \frac{2^2\cdot 5}{3^2\cdot 7\cdot 11}\\
P_4^2P_6 &: (P_4P_4)_6P_6 = \frac{2^2\cdot 5}{3^2\cdot 11}P_6^2 & \to \frac{2^2\cdot 5}{3^2\cdot 11\cdot 13}\\  
P_2P_6^2 &: P_2(P_6P_6)_2 = \frac{2\cdot 7}{11\cdot 13}P_2^2 & \to \frac{2\cdot 7}{5\cdot 11\cdot 13}\\
P_2P_4P_6 &: (P_2P_4)_6\:P_6 = \frac{5}{11}P_6^2 & \to  \frac{5}{11\cdot 13}\\
P_2P_4P_8 &: (P_2P_4)_8\:P_8 & \to 0\quad .
\end{array}\end{equation}
Similarly in $\Sigma^2$, terms $P_nP_m$ yield a factor $1/(2n+1)$ if $n=m$ and 0 otherwise, so that
\[
	\Sigma^2 = s_0^2 \:+\: \frac{1}{5}\,s_2^2 \:+\: \frac{1}{3^2}\,s_4^2 \:+\: \frac{1}{13}\,s_6^2 \quad.
\]
Out of $\Sigma^1$ in the integral in Eq.~(\ref{eq:s0integral1}), only $s_0P_0$ survives.  
Now, $s_0$ can be expressed in terms of the $s_{n,n\geq 2}$, so that 
\[
	s_0:=s_0^{(1)} + s_0^{(2)} + \ldots + s_0^{(7)}, 
\]
where $O$ denotes the order of expansion. Sorting terms according to their order, we obtain (note the $-$sign)
\begin{eqnarray} 
- s_0^{(1)} &=& 0\nonumber\\
- s_0^{(2)} &=& \frac{1}{5}\: s_2^2\nonumber\\ 
- s_0^{(3)} &=& \frac{2}{3\cdot 5\cdot 7}\: s_2^3\nonumber\\
- s_0^{(4)} &=&  \frac{1}{3^2}\: s_4^2 \:+\: \frac{2}{5\cdot 7}\: s_2^2 s_4\nonumber\\
- s_0^{(5)} &=& \frac{2}{3\cdot 5^2\cdot 7}\:s_2^5 \:+\: \frac{2^2\cdot 5}{3^2\cdot 7\cdot 11}\: s_2 s_4^2\nonumber\\
- s_0^{(6)} &=&  \frac{-127}{3^2\cdot 5^3\cdot 7^2}\:s_2^6 \:+\:  \frac{2}{5^2\cdot 7}\:s_2^4 s_4 
	+ \frac{2\cdot 3}{7\cdot 11\cdot 13}\: s_4^3 \:+\: \frac{2\cdot 5}{11\cdot 13}\:s_2 s_4 s_6 
	\:+\: \frac{1}{13}\:s_6^2\nonumber\\
- s_0^{(7)} &=& \frac{2}{3\cdot 5^3\cdot 7}\:s_2^7 \:+\: \frac{2\cdot 41}{3^3\cdot 5\cdot 7\cdot 11}\:s_2^3 s_4^2
	\:+\: \frac{2^3}{3\cdot 5^2\cdot 7}\:s_2^5 s_4 \:+\: \frac{2\cdot 7}{ 5\cdot 11\cdot 13}\: s_2 s_6^2 
	\:+\: \frac{2^2 \cdot 5}{3^2\cdot 11\cdot 13}\: s_4^2 s_6\:. \nonumber\\
	&& \quad
\end{eqnarray}

\bibliography{ms_tof7}
\bibliographystyle{aasjournal}
\end{document}